\documentclass[]{pasj01}
\Received{2023/05/31}%{yyyy/mm/dd}
\Accepted{2023/08/02}%{yyyy/mm/dd}
\usepackage{comment}
\usepackage{ulem}

\begin{document}

\title{\textit{N}-body simulation of planetary formation through pebble accretion in a radially structured protoplanetary disk}
\author{Tenri \textsc{Jinno},\altaffilmark{1,}$^{*}$
	Takayuki R. \textsc{Saitoh},\altaffilmark{1,}$^{*}$
        Yota \textsc{Ishigaki},\altaffilmark{2,}$^{*}$ and
	Junichiro  \textsc{Makino}\altaffilmark{1,}$^{*}$}
\altaffiltext{1}{Department of Planetology, Graduate School of Science, Kobe University, 1-1 Rokkodai-cho, Nada-ku, Kobe, Hyogo 657-8501, Japan}
\altaffiltext{2}{Department of Solar System Science, Institute of Space and Astronautical Science, Japan Aerospace Exploration Agency}
\email{223s415s@gsuite.kobe-u.ac.jp, saitoh@people.kobe-u.ac.jp, ishigaki@alumni.u-tokyo.ac.jp, jmakino@people.kobe-u.ac.jp}

\maketitle
\begin{abstract}
In the conventional theory of planet formation, it is assumed that protoplanetary disks are axisymmetric and have a smooth radial profile. However, recent radio observations of protoplanetary disks have revealed that many of them have complex radial structures. In this study, we perform a series of \textit{N}-body simulations to investigate how planets are formed in protoplanetary disks with radial structures. For this purpose, we consider the effect of continuous pebble accretion onto the discontinuity boundary within the terrestrial planet-forming region ($\sim0.6$ AU). We found that protoplanets grow efficiently at the discontinuity boundary, reaching the Earth mass within $\sim10^4$ years. We confirmed that giant collisions of protoplanets occur universally in our model. Moreover, we found that multiple planet-sized bodies form at regular intervals in the vicinity of the discontinuity boundary. These results indicate the possibility of the formation of solar system-like planetary systems in radially structured protoplanetary disks.
\end{abstract}

\section{Introduction} \label{sec:Introduction}
One of the pioneering studies of planetary formation theory was conducted by Chushiro Hayashi's group at Kyoto University \citep{1981PThPS..70...35H,1985prpl.conf.1100H}. In their scenario, the planetary formation process starts from a disk that is stable against self-gravity. This disk is assumed to have a dust component consistent with the current mass of planets, and the dust-gas ratio comes from the solar abundance (the Minimum-Mass Solar Nebula model, MMSN). First,  dust condenses to the midplane of the disk. Then, planetesimals are formed through the gravitational instability of dust. The planetesimals grow through mutual collisions to form planets. Furthermore, they showed that when the mass of a protoplanet reaches a certain critical value ($\simeq 10 M_{\oplus}$), disk gas flows into the planet, forming a gas giant \citep{10.1143/PTP.64.544}. Since this model successfully described the sequence of terrestrial planets, gas giants, and icy planets, it came to be regarded as the standard theory of planet formation, even though many difficulties were known.

Some of the most critical problems have been resolved. For example, the problem of the formation time of Jupiter has been resolved by the realization that planetesimals grow through the runaway process \citep{1989Icar...77..330W,1993Icar..106..210I,1996Icar..123..180K}. However, several problems remain unresolved, including planet migration, the occurrence of magneto-rotational instability, and the diversity of observed protoplanetary disks. In the following, we outline these problems. \par

First, it has been found that protoplanets that have grown to the size of Mars experience the Type I migration and drift toward the Sun due to gravitational interaction with the gas disk \citep{1986Icar...67..164W,2002ApJ...565.1257T}. In the MMSN disk, the Type I migration timescale of Mars-size planets at 1 AU is estimated as $1\times10^6$ yr. The orbital distribution of the planets after migration due to the Type I effect cannot reproduce the orbital distribution of the solar system \citep{2008ApJ...673..487I}.
 
Second, the occurrence of the magneto-rotational instabilities (MRI) in accretion disks was found in the early 1990s \citep{1991ApJ...376..214B}, indicating that the structure of protoplanetary disks is much more complex than what was assumed in the standard theory. More recently, it has been pointed out that in many protoplanetary disks, turbulence originating from MRI does not occur or is suppressed in certain radial ranges \citep{2013ApJ...769...76B}. This region is called the dead zone. This dead zone is considered to be a laminar structure sandwiched by two turbulent outer layers \citep{2011ARA&A..49..195A}, since cosmic rays can keep the ionization ratio of the outer layers high enough for MRI to occur. In this turbulent region, the existence of magnetic disk winds has been theoretically predicted. This disk wind is expected to be the mechanism to remove the angular momentum from protoplanetary disks \citep{2016A&A...596A..74S}. Planetary system formation in protoplanetary disks with dead zone and magnetic disk winds is now being actively studied (e.g., \cite{2009A&A...497..869L,2021A&A...650A.116M}). 

Third, recent observations using Atacama Large Millimeter/sub-millimeter Array (ALMA) have revealed the presence of non-axisymmetric dust distributions and ring-like structures within protoplanetary disks (e.g., \cite{2015PASJ...67..122M,2018ApJ...869L..41A}). The standard theory, which assumes a smooth disk structure, cannot explain these sub-structures observed in protoplanetary disks or the planet formation process within the structured disk.

In recent years, pebble accretion has attracted attention (\authorcite{2012A&A...539A.148B} \yearcite{2012A&A...539A.148B}; \authorcite{2012A&A...544A..32L} \yearcite{2012A&A...544A..32L,2014A&A...572A.107L}). Smaller pebbles, typically on the order of cm-size, are instrumental in driving pebble accretion. They efficiently migrate inward through hydrodynamic drag \citep{1977MNRAS.180...57W,1986Icar...67..375N}. These pebbles may accumulate at the boundary between the dead zone and the inner turbulent zone where a pressure bump exists \citep{2010ApJ...714.1155K}. In addition, it has been pointed out that in the dead zone, the radial velocity of pebbles may be slowed down by pebble-gas back-reaction, potentially resulting in a runaway pile-up of pebbles (\cite{2021A&A...645L...9H};  \yearcite{2022A&A...660A.117H}). In either case, the supply of pebbles to the core feeding region may lead to efficient planet formation.

As we have seen above, the standard theory has many problems, and many ideas have been proposed to solve them. In this study, we constructed a disk model that incorporates ideas to overcome these problems in a consistent manner as a whole and performed \textit{N}-body simulations to study the formation of terrestrial planets. It should be noted that, even though we start with a specific disk model inspired by that of \citet{2017NewA...54....7E} (hereafter EI17), our scenario can be applied to any disk model which has the inner cutoff at around 1AU (e.g., \cite{2016A&A...596A..74S}).

\textit{N}-body simulations in which the pebble accretion process to planetary cores is taken into account have been performed in several studies (e.g., \authorcite{2017A&A...607A..67M} \yearcite{2017A&A...607A..67M,2021A&A...650A.116M}, \cite{2022A&A...668A.170L} and \cite{2023MNRAS.518.3877J}). The first two studies have simulated the growth of planetary embryos, assuming that they grow through pebble accretion. The growth rate of embryos is given by the analytic function. This implies that in their model how and where the embryos are formed and how they grow are model assumptions. The latter two studies have investigated the formation and evolution of planetesimals, taking into account pebble accretion in the rings induced by permanent pressure bumps. They determined the locations of the pressure bump, motivated by the observed ring structures (e.g., \cite{2018ApJ...869L..41A}). Therefore, both studies focus on planet formation at pressure bumps located at distances of 10 AU and 75 AU from the central star, which are outside the region where terrestrial planets form. 

In addition to the studies discussed above, \citet{2016MNRAS.460.2779C} have investigated the process of gas giant formation within protoplanetary disks that contain multiple ring-like structures using \textit{N}-body simulations. They include spatially and temporally varying viscous stresses within a limited radius, allowing for the reproduction of the formation and decay of pressure-induced rings within the disk. However, it should be noted that their research does not encompass the formation of pressure-induced rings within the region where terrestrial planet formation occurs. There have been studies that utilize planet population synthesis calculations that account for planetary growth by core accretion and planetary migration by Type-I and Type-II migrations (e.g., \cite{2018MNRAS.478.2599A}).

We intend to study the formation and growth of terrestrial planets from infalling pebbles, which are trapped at the inner boundary of the dead zone ($\sim0.6$ AU). We model pebbles as $N$-body particles and solve their interactions directly through large-scale $N$-body simulations with up to 1 million particles.\par

This paper is organized as follows. In section \ref{sec:Method}, we present our model for the structure of the protoplanetary disk, the dust growth, and the growth of planets. The last part is modeled by $N$-body simulation.  In section \ref{sec:Results}, we present the results of \textit{N}-body simulations and show how pebbles grow to planets size at the inner boundary of the disk. We also show the results of \textit{N}-body simulations on disk models with various gas-dust ratios and particle numbers. In section \ref{sec:Discussion}, we overview the growth scenario of planets in our model and compare it with other models. In section \ref{sec:Summary}, we make concluding remarks.

\section{Model and Numerical Method} \label{sec:Method}
\subsection{Overview of Our Model} \label{sec:Method:overview}
The aim of our study is to investigate how planet formation proceeds when we take into account the continuous pebble accretion to discontinuous boundaries of the protoplanetary disk. To do so, we constructed a protoplanetary disk model with a discontinuous structure and a model for the growth and migration of dust particles in it. By combining these two models, we obtain the time-varying mass inflow rate at the disk inner boundary. We conducted a series of $N$-body simulations of planetary growth under this inflow rate.\par

We constructed a gas disk model following EI17. The outer and inner regions of the disk are turbulent while the intermediate region is non-turbulent (i.e., the dead zone). Such a disk with 
a dead zone seems to be the natural outcome of the theoretical model of protoplanetary disks \citep{2011ARA&A..49..195A}. In such a disk, a pressure bump forms at the boundary of the dead zone, causing dust particles to accumulate \citep{2010ApJ...714.1155K,2014ApJ...780...53C}.
For simplicity, we assume that our gas disk is a stationary one-dimensional $\alpha$ disk \citep{1973A&A....24..337S}.\par

We also constructed a model for the growth and migration of the dust particles in the gas disk to determine the mass inflow rate at the inner boundary of the dead zone \citep{2017AREPS..45..359J}. This is how we express pebble accretion. \par

Using these models, we determine the mass inflow rate at the inner boundary of the dead zone. We construct the initial condition for our $N$-body simulation to realize this inflow rate (see section \ref{sec:Method:numerical} for details). We conducted $N$-body simulations with initial particle numbers ranging from $6\times10^{4}$ to $10^{6}$.

\subsection{The gas disk} \label{sec:Method:gasdisk}
Here we present our protoplanetary disk model surrounding a solar mass star.

\subsubsection{The surface density of the gas disk}\label{sec:Method:gasdisk:surface}
Our disk model is one-dimensional. The time evolution of the surface density of the gas disk $\Sigma$ is given by the continuity equation
\begin{equation}
\frac{\partial{\Sigma}}{\partial{t}}=\frac{1}{2\pi r}\frac{\partial{\dot{M}}}{\partial{r}}, \label{eq.1}
\end{equation}
where the mass accretion rate $\dot{M}$ (positive for inward accretion) is defined using the gas radial velocity $v_{r}$ as:
\begin{equation}
\dot{M}=-2\pi r\Sigma v_r. \label{eq.2}
\end{equation}
We neglect the time evolution of the disk and regard it as in a steady state. In this case, we can rewrite eq.(\ref{eq.1}) as
\begin{equation}
\frac{\partial{\dot{M}}}{\partial{r}}=0. \label{eq.3}
\end{equation}
Using eqs.(\ref{eq.2}), (\ref{eq.3}), and the azimuthal direction component of the fluid equation of motion
\begin{equation}
\Sigma v_r r\frac{\partial}{\partial r}(r^2\Omega)=\frac{\partial}{\partial r}\left(r^3\Sigma\nu\frac{\partial\Omega}{\partial r}\right), \label{eq.4}
\end{equation}
 we obtain
\begin{eqnarray}
\dot{M}&=&-2\pi r\Sigma v_r,  \nonumber \\
&=& 6\pi r^{1/2}\frac{\partial}{\partial r} \left(\Sigma\nu r^{1/2}\right)=\mathrm{const}, \label{eq.5}
\end{eqnarray}
where $\nu$ and $\Omega$ are  the disk viscosity and the Keplerian orbital frequency. Here $\Omega$ is given as
\begin{equation}
\Omega=\left(\frac{GM_{*}}{r^3}\right)^{1/2}=2.0\times10^{-7}\left(\frac{r}{\mathrm{AU}}\right)^{-3/2} ~~\mathrm{s}^{-1}, \label{eq.8}
\end{equation}
where $G$ is the gravitational constant and $M_{*}$ is the mass of the central star. We set $M_{*}=1M_{\odot}$.
Integrating eq.(\ref{eq.5}) with respect to $r$, we obtain
\begin{equation}
\Sigma = \frac{\dot{M}}{3\pi \nu}. \label{eq.14}
\end{equation}
We assume that the disk is an $\alpha$-disk so that the viscosity can be expressed by
\begin{equation}
\nu = \alpha c_{\mathrm{s}}H, \label{eq.6}
\end{equation}
where $\alpha$ is the viscous parameter, $c_{\mathrm{s}}$ is the sound velocity, and $H$ is the scale height of the disk. The functional forms of $c_{\mathrm{s}}$ and $H$ are given as follows:
\begin{eqnarray}
c_{\mathrm{s}}&=&\left(\frac{k_{\mathrm{B}}T_{\mathrm{m}}}{\mu m_{\mathrm{H}}}\right)^{1/2}=1.0\times10^5\left(\frac{T_{\mathrm{m}}}{280\mathrm~{K}}\right)^{1/2}~~\mathrm{cm}~\mathrm{s}^{-1}, \label{eq.7} \\
H&=&\frac{c_{\mathrm{s}}}{\Omega}=5.0\times10^{11}\left(\frac{T_{\mathrm{m}}}{280\mathrm~{K}}\right)^{1/2}\left(\frac{r}{\mathrm{AU}}\right)^{3/2}~~\mathrm{cm}, \label{eq.9}
\end{eqnarray}
where $T_{\mathrm{m}}$ is the mid-plane temperature of the gas disk (see section \ref{sec:Method:gasdisk:temperature}), $k_{\mathrm{B}}$ is the Boltzmann constant, $\mu=2.34$ is the mean molecular weight of the gas and $m_{\mathrm{H}}$ is the mass of a hydrogen atom.\par

The value of the viscous parameter $\alpha$ is different for turbulent and non-turbulent regions \citep{2020apfs.book.....A}. Here we adopt the following values
\begin{equation}
\alpha =\left\lbrace  
			\begin{array}{ll}
                    \alpha_{\mathrm{act}}=1.0\times10^{-2} \\ \mathrm{for~turbulent~region}\\
                    \alpha_{\mathrm{inact}}=\gamma\alpha_{\mathrm{act}}=1.0\times10^{-2.5} \\ \mathrm{for~non\mathchar`-turbulent~region}.
			\end{array}
		\right. \label{eq.10}
\end{equation}
where $\alpha_{\mathrm{act}}$ is determined using the result of numerical MHD simulations of planetary disks \citep{2010ApJ...713...52D}. We assumed the reduction factor of turbulent $\gamma$ to be $\gamma=10^{-0.5}\simeq0.316$ for the entire non-turbulent region, though $\gamma$ has not been constrained well yet. \par

EI17 solved the steady-state equation of the disk to obtain the locations of the inner and outer boundaries of the dead zone. Here, we adopt their numerical results to determine the location of the boundaries as a function of the mass accretion rate $\dot{M}$. We employed single power law forms to fit the numerical results and these are:
\begin{eqnarray}
r_{\mathrm{in}}&=&1.1\times10^{3}\dot{M}^{0.47}~~~\mathrm{AU}, \label{eq.11} \\
r_{\mathrm{out}}&=&6.6\times10^{5}\dot{M}^{0.62}~~~\mathrm{AU}, \label{eq.12}
\end{eqnarray}
where $r_{\mathrm{in}}$ and $r_{\mathrm{out}}$ represent the inner and outer boundaries of the dead zone, respectively. The region between $r_{\mathrm{in}}$ and $r_{\mathrm{out}}$ is MRI inactive and $\alpha=\alpha_{\mathrm{inact}}$ for this region. Other regions are MRI active and $\alpha=\alpha_{\mathrm{act}}$ for these regions. These coefficients and power law indexes are determined to reproduce EI17 results: $r_{\mathrm{in}}=1$ AU and $r_{\mathrm{out}}=60$ AU for $\dot{M}=10^{-6.5}~M_{\odot}/$yr and $r_{\mathrm{in}}=0.2$ AU and $r_{\mathrm{out}}=7$ AU for $\dot{M}=10^{-8.0}~M_{\odot}/$yr.\par

\subsubsection{The mid-plane temperature of the gas disk} \label{sec:Method:gasdisk:temperature}
In order to determine $\Sigma$, we need the temperature of the disk. Following \citet{1990ApJ...351..632H}, we use the following equation for 
the mid-plane temperature $T_{\mathrm{m}}$:  
\begin{equation}
T_{\mathrm{m}}^4=\left( \frac{3\dot{M}\Omega^2}{8\pi \sigma} \right) 
\left(\frac{3}{8}\frac{\alpha}{\alpha_{\mathrm{act}}}\frac{\kappa\Sigma}{2}+\frac{\sqrt{3}}{4} \right)+T_{\mathrm{irr}}^4, \label{eq.15}
\end{equation}
where $\sigma$ is the Stefan-Boltzmann constant, $\alpha_{\mathrm{act}}$ is the viscous parameter for the MRI active state, $\kappa$ is the opacity of the gas disk, and $T_{\mathrm{irr}}$ is the irradiation temperature due to the central star. We adopt the opacity of the gas disk $\kappa$  (in unit of cm$^2$/g) from \citet{1998Icar..132..100S}:
\begin{eqnarray}
\kappa=
        \left\{\begin{array}{ll}
            2\times10^{-4}T_{\mathrm{m}}^2 &T_{\mathrm{m}}<150~\mathrm{K}\\
            1.15\times10^{18}T_{\mathrm{m}}^{-8} &150~\mathrm{K}\leq T_{\mathrm{m}}<180~\mathrm{K}\\
            2.13\times10^{-2}T_{\mathrm{m}}^{3/4} &180~\mathrm{K}\leq T_{\mathrm{m}}<1380~\mathrm{K}\\
            4.38\times10^{44}T_{\mathrm{m}}^{-14} &T_{\mathrm{m}}\geq 1380~\mathrm{K}
        \end{array}\right. \label{kappa}
\end{eqnarray}
The irradiation temperature $T_{\mathrm{irr}}$ is given by \citet{10.1093/mnras/stu1715}:
\begin{equation}
T_{\mathrm{irr}}^4=\frac{1}{2}(1-\epsilon)T_{*}^4 \left( \frac{R_{*}}{r} \right)^2\left[\frac{4}{3\pi} \left(\frac{R_{*}}{r}\right)+\frac{2}{7}\frac{H}{r} \right], \label{eq.16}
\end{equation}
where $\epsilon$ is the albedo of the disk, $T_{*}$ and $R_{*}$ are the temperature and radius of the central star. Following \citet{10.1093/mnras/stu1715}, we set values of $\epsilon$, $T_{*}$ and $R_{*}$ to be 0.5, 4000 K and 3$R_{\odot}$ respectively. 

\subsubsection{The overall structure of the gas disk}\label{sec:Method:gasdisk:detailed}
%Figure. 2
\begin{figure}[hbtp]
 \includegraphics[width= 8cm,height=6.5cm]{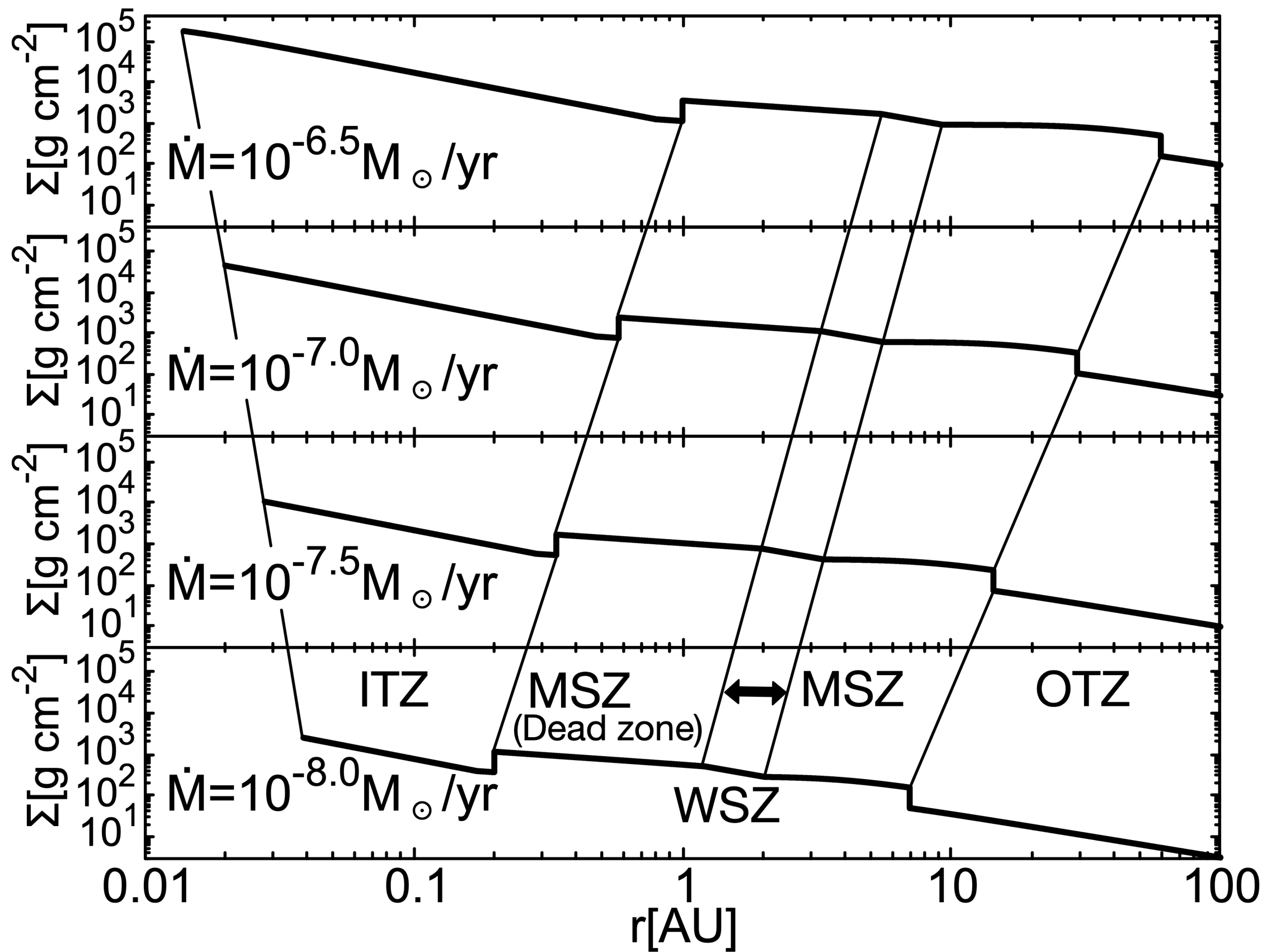}
 \caption{The radial profile of the surface density of the gas disk with different mass accretion rates ($\dot{M}=10^{-8.0} - 10^{-6.5}~M_{\odot}/\mathrm{yr}$).}
 \label{fig:disk_structure}
\end{figure}
Fig.\ref{fig:disk_structure} shows the surface density profile of the gas disk for four different mass accretion rates ranging from $10^{-8.0}~M_{\odot}/\mathrm{yr}$ to $10^{-6.5}~M_{\odot}/\mathrm{yr}$.
For all cases, there are two discontinuous boundaries in the surface density. We call them inner dead zone boundary (IDB) and outer dead zone boundary (ODB). As we have discussed in section \ref{sec:Method:gasdisk:surface}, we used the result of the 1D disk calculation of EI17 to determine the location of these boundaries. \par

We call the inner and outer turbulent zones inner turbulent zone (ITZ) and outer turbulent zone (OTZ) respectively. The non-turbulent zone, which we call the MRI-suppressed zone (MSZ) exists between ITZ and OTZ. Slight slope changes are seen in MSZ. This reflects the change of opacity in the gas disk due to the sublimation of ice at temperature 180--150 K [eq.(\ref{kappa})]. We call this region the water sublimation zone (WSZ).\par

Near the central star, the gas disk is truncated by the magnetic field of the central star. The radius of disconnection is called the Alfv\'{e}n radius $r_{\mathrm{A}}$, which is the radius where the magnetic pressure of the central star is balanced by the gas pressure of the accreting gas disk. Since $r_{\mathrm{A}}$ is proportional to $\dot{M}^{-2/7}$, $r_{\mathrm{A}}$ moves outward as the mass accretion rate decreases.

The region of interest in this study is the vicinity of IDB, where a discontinuity in the gas surface density profile is created by the change in the value of $\alpha$.  This surface density gap creates a pressure bump in IDB. It is expected that planets form efficiently as dust that accretes from the outer regions gets trapped near IDB due to the pressure bump \citep{2010ApJ...714.1155K,2014ApJ...780...53C}.

\subsection{Evolution of dust particles in the gas disk} \label{sec:Method:ParticleGrowth}
In this subsection, we first present the physical background of the growth and migration of the dust particles and describe the assumption made in our model (section \ref{sec:Method:ParticleGrowth:model_assumption}). Then, in section \ref{sec:Method:ParticleGrowth:mass} and \ref{sec:Method:ParticleGrowth:dust}, we describe our model for the growth and migration of the dust particle within the disk presented in the previous subsection. 

\subsubsection{Physical background and modeling assumption of dust particles growth and migration} \label{sec:Method:ParticleGrowth:model_assumption}
Dust particles outside of IDB ($r>r_{\mathrm{in}}$) grow via collisions with other dust particles. As dust particles continue to grow through mutual collisions, they drift inward due to the effect of the gas drag \citep{1976PThPh..56.1756A}. This inward migration continues until dust particles reach IDB, where the presence of pressure bump halts their inward migration \citep{2010ApJ...714.1155K,2014ApJ...780...53C}.\par
We solve the growth and migration of dust particles simultaneously to determine the mass inflow rate onto IDB. The radial migration timescale for dust particles ranging from $\mu$m to $\sim$cm in size (i.e., with a mass of $\sim10^4$ g or less) is significantly longer than the growth timescale (see section \ref{sec:Method:ParticleGrowth:dust}). Thus, in our model, we initially approximate that dust particles grow through mutual collisions in situ, and once they have grown to critical mass $m_{\mathrm{c}}$, they subsequently drift towards the Sun due to the effect of the gas drag. This approximation is similar to the idea proposed by \citet{2014A&A...572A.107L}. They proposed a pebble flux calculation model based on the idea that all the dust grows to the drift limits at a certain radius, decoupled from the gas, and drifts inwards.

Our approximation for dust particle growth and migration allows us to calculate the growth of dust particles and the drift of particles separately. Thus, we can define the mass accretion timescale onto IDB, $t_{\mathrm{acc}}$, as the sum of the time it takes for a dust particle to grow to $m_{\mathrm{c}}$ and its radial migration timescale (see Appendix \ref{sec:Appendix_mass_flux} for details).
\subsubsection{The growth rate of dust particles} \label{sec:Method:ParticleGrowth:mass}
In our model, dust particles grow via collisions with other dust particles. The rate of dust particle growth is given by
\begin{equation}
\frac{d m_{\mathrm{p}}}{dt}=\dot{m}_{\mathrm{p}}=\pi a_{\mathrm{p}}^2\rho_{\mathrm{p}}v_{\mathrm{rel,pp}}\left(1+\frac{2Gm_{\mathrm{p}}}{a_{\mathrm{p}}v_{\mathrm{rel,pp}}^2} \right), \label{eq.17}
\end{equation}
where $m_{\mathrm{p}}$ is the particle mass, $a_{\mathrm{p}}$ is the dust particle radius, $\rho_{\mathrm{p}}$ is the dust particle density at the particle scale height, and $v_{\mathrm{rel,pp}}$ is the particle-particle relative velocity. By integrating eq.(\ref{eq.17}), we determine the evolution of $m_{\mathrm{p}}$. In the following, we describe how we determine the parameters in  eq.(\ref{eq.17}).\par
The particle radius $a_{\mathrm{p}}$ in eq.(\ref{eq.17}) is given by
\begin{equation}
a_{\mathrm{p}}=\left(\frac{3m_{\mathrm{p}}}{4\pi \rho_{\mathrm{i}}} \right)^{1/3}. \label{eq.18}
\end{equation}
Since we consider the dust particles inside WSZ, here we use the dust particle internal density $\rho_{\mathrm{i}} = 2~\mathrm{g~cm^{-3}}$ which is appropriate for carbonaceous material.\par
The dust particle density at the particle scale height is given by
\begin{equation}
\rho_{\mathrm{p}}=\frac{\bar{f}\Sigma}{\sqrt{2\pi}z_{\mathrm{p}}}\exp\left(-\frac{1}{2}\right), \label{eq.19}
\end{equation}
where $\bar{f}$ is the dust-to-gas fraction and $z_{p}$ is the scale height of the dust particles. Here, we adopt $\bar{f}=\bar{f}_{\mathrm{MMSN}}=2.5\times10^{-3}$ (for IDB $<r<$ WSZ) where $\bar{f}=\bar{f}_{\mathrm{MMSN}}$ is what assumed in \citet{1985prpl.conf.1100H}.\par
The scale height of dust particles $z_{\mathrm{p}}$ is given by EI17 as:
\begin{eqnarray}
\frac{d z_{\mathrm{p}}}{dt}=-v_{z{\mathrm{p}}}~~~& &\mathrm{for}~z_{\mathrm{p}}>H_{\mathrm{p}}, \label{eq.27}\\
z_{\mathrm{p}}=H_{\mathrm{p}} ~~~& &\mathrm{for}~z_{\mathrm{p}}<H_{\mathrm{p}}, \label{eq.28}
\end{eqnarray}
 where $H_{\mathrm{p}}$ and $v_{z{\mathrm{p}}}$ are the particle scale height in equilibrium and the particle settling velocity. In our model, we adopt the scale height of the gas disk $H$ as the initial value of $z_p$. The particle scale height in equilibrium $H_{\mathrm{p}}$ is given by \citet{2007ApJ...662..613Y}: 
\begin{equation}
H_{\mathrm{p}}=\left(1+\frac{\Omega t_{\mathrm{s}}}{\alpha_{\mathrm{D}}} \right)^{-\frac{1}{2}}\left(1+\frac{\Omega t_{\mathrm{s}}}{\Omega t_{\mathrm{s}}+1} \right)^{-\frac{1}{2}}H, \label{eq.20}
\end{equation}
 where $\Omega$ is the Keplerian orbital frequency given by eq.(\ref{eq.8}), $t_{\mathrm{s}}$ is the particle stopping time, and $\alpha_{\mathrm{D}}$ is the effective viscous parameter of the gas disk in the region where dust particles exist.
Following \citet{2016A&A...589A..15S}, we use the equation for the particle stopping time $t_{\mathrm{s}}$:
\begin{equation}
t_{\mathrm{s}}=\left\{
\begin{array}{ll}
\frac{4\rho_{\mathrm{i}}a_{\mathrm{p}}^2}{9\rho_{\mathrm{g}}v_{\mathrm{th}}\lambda} &\mathrm{for}~(a_{\mathrm{p}}>\frac{9}{4}\lambda), \\
\frac{\rho_{\mathrm{i}}a_{\mathrm{p}}}{\rho_{\mathrm{g}}v_{\mathrm{th}}} & \mathrm{for}~(a_{\mathrm{p}}\leq\frac{9}{4}\lambda), \\
\end{array}\right.
\end{equation}
 where $\rho_{\mathrm{g}}, v_{\mathrm{th}}$ and $\lambda$ are the gas density, the thermal velocity, and the mean free path of the gas molecule at given $r$ and $z$. These parameters are given as follows:
\begin{eqnarray}
\rho_{\mathrm{g}}&=&\frac{\Sigma}{\sqrt{2\pi}H}\exp\left(-\frac{z^2}{2H^2}\right), \label{eq.22} \\
v_{\mathrm{th}}&=&\sqrt{\frac{8k_{\mathrm{B}}T_{\mathrm{m}}}{\pi\mu}}, \label{eq.23}\\
\lambda &=& \frac{\mu}{{\sigma_{\mathrm{col}}\rho_{\mathrm{g}}}}. \label{eq.24}
\end{eqnarray}
Here, $\sigma_{\mathrm{col}}=2.0\times 10^{-15}$cm$^2$ is the molecular collision cross-section. The particle stopping time is expressed in terms of the dimensionless Stokes number as
 \begin{equation}
 \mathrm{St}\equiv \Omega t_{\mathrm{s}}. \label{eq.25}
 \end{equation}
 We, hereafter use St instead of $\Omega t_{\mathrm{s}}$.
The particle settling velocity in eq.(\ref{eq.27}) is given by 
 \begin{equation}
v_{z{\mathrm{p}}}=\frac{\mathrm{St}}{\mathrm{St}+1}z_{\mathrm{p}}\Omega. \label{eq.29}
 \end{equation}

The particle-particle relative velocity is given by \citet{2007A&A...466..413O} as:
\begin{eqnarray}
v_{\mathrm{rel,pp}}&=&\sqrt{v_{\mathrm{B}}^2+v_{r\mathrm{pp}}^2+v_{\phi\mathrm{ pp}}^2+v_{z\mathrm{pp}}^2+v_{\mathrm{turb,pp}}^2}, \label{eq.30}\\
v_{\mathrm{B}}&=&\sqrt{\frac{16}{\pi}\frac{k_{\mathrm{B}}T_{\mathrm{m}}}{m_{\mathrm{p}}}}, \label{eq.31} \\
 v_{r\mathrm{pp}}&=&\left(\frac{2\mathrm{St}}{1+(\mathrm{St})^2}-\frac{\mathrm{St}}{1+(0.5\mathrm{St})^2}\right)\eta r\Omega, \label{eq.32}\\
  v_{\phi\mathrm{pp}}&=&-\left(\frac{(\mathrm{St})^2}{1+(\mathrm{St})^2}-\frac{(0.5\mathrm{St})^2}{1+(0.5\mathrm{St})^2}\right)\eta r\Omega, \label{eq.33} \\
 v_{z\mathrm{pp}}&=&\left(\frac{\mathrm{St}}{1+\mathrm{St}}-\frac{0.5\mathrm{St}}{1+0.5\mathrm{St}}\right)z_{\mathrm{p}}\Omega, \label{eq.34} 
\end{eqnarray}
\begin{eqnarray}
v_{\mathrm{turb,pp}}=\nonumber\\& \hspace{-1.3cm} \sqrt{\alpha_{\mathrm{D}}}c_{\mathrm{s}}\times   \left \{
\begin{array}{ll}
\mathrm{Re}^{1/4}\Omega|0.5t_{\mathrm{s}}| &\mathrm{for~}\mathrm{St}<8\mathrm{Re}^{-1/2}\\
\sqrt{2\mathrm{St}} &\mathrm{for~}8\mathrm{Re}^{-1/2}\leq\mathrm{St}<1\\
\sqrt{\frac{1}{1+\mathrm{St}}+\frac{1}{1+0.5\mathrm{St}}} &\mathrm{for~}						1\leq\mathrm{St},
\end{array} \right. \label{eq.35}
\end{eqnarray}
where $v_{\mathrm{B}}, v_{r\mathrm{pp}}, v_{\phi\mathrm{pp}}, v_{z\mathrm{pp}}$ and $v_{\mathrm{turb,pp}}$ are Brownian motion, radial drift difference, azimuthal drift difference, vertical settling difference, and turbulent velocity, respectively. Here, we assume that the relative velocity between particles can be represented by the velocity difference between two particles with $t_{\mathrm{s}}$ and $0.5t_{\mathrm{s}}$, following \citet{2016A&A...589A..15S}. Following EI17, we use the effective viscous parameter:
 \begin{equation}
\alpha_{\mathrm{D}}=\left\{
\begin{array}{ll}
0 & \mathrm{for}~\rho_{\mathrm{pm}}<\rho_{\mathrm{gm}} \\
0.19\left(\frac{\eta r}{H}\right)\min(\mathrm{St},1) & \mathrm{for}~\rho_{\mathrm{pm}}>\rho_{\mathrm{gm}}, \\
\end{array}\right. \label{eq.26}
\end{equation}
 where $\eta$ is a dimensionless quantity that characterizes the pressure gradient of the gas disk and $\rho_{\mathrm{pm}}, ~\rho_{\mathrm{gm}}$ are the dust and gas volume densities at the mid-plane of the disk. Here, $\eta$, $\rho_{\mathrm{pm}}$ and $\rho_{\mathrm{gm}}$ are given by
\begin{eqnarray}
\eta&=&-\frac{1}{2}\frac{c_{\mathrm{s}}^2}{r^2\Omega^2}\left(\frac{\partial \log(\rho_{\mathrm{gm}}T_{\mathrm{m}})}{\partial \log r} \right), \label{eta}\\
\rho_{\mathrm{pm}}&=&\frac{\bar{f}\Sigma}{\sqrt{2\pi}z_{\mathrm{p}}},\label{rho_pm} \\
\rho_{\mathrm{gm}}&=&\frac{\Sigma}{\sqrt{2\pi}H}.\label{rho_m}
\end{eqnarray}
We adopt the turbulent Reynolds number Re from EI17:
\begin{equation}
\mathrm{Re}=\frac{2\alpha_{\mathrm{D}}c_{\mathrm{s}}^2}{\Omega\lambda v_{\mathrm{th}}}. \label{eq.36}
\end{equation}
In our dust growth model, we do not consider the viscous stirring velocity $v_{\mathrm{VS}}$ considered in EI17 because it is orders of magnitude smaller than other velocity components when the dust is $\mathrm{\mu m}$ to cm size.\par
Using eq.(\ref{eq.18}) to eq.(\ref{eq.36}), we are able to integrate the ordinary differential eq.(\ref{eq.17}) by the Euler method. The time steps for the integration are:
\begin{equation}
\Delta t= \left\{\begin{array}{ll}
        10^{-1} \mathrm{~yr~~~for~} m_{\mathrm{p}}>10^{-4}\mathrm{~g}\\
        10^{-3} \mathrm{~yr~~~for~} m_{\mathrm{p}}<10^{-4}\mathrm{~g}. \label{eq.37}
        \end{array}\right.
\end{equation}\par
\begin{figure}[hbtp]
 %\centering
 \includegraphics[width= 8cm]{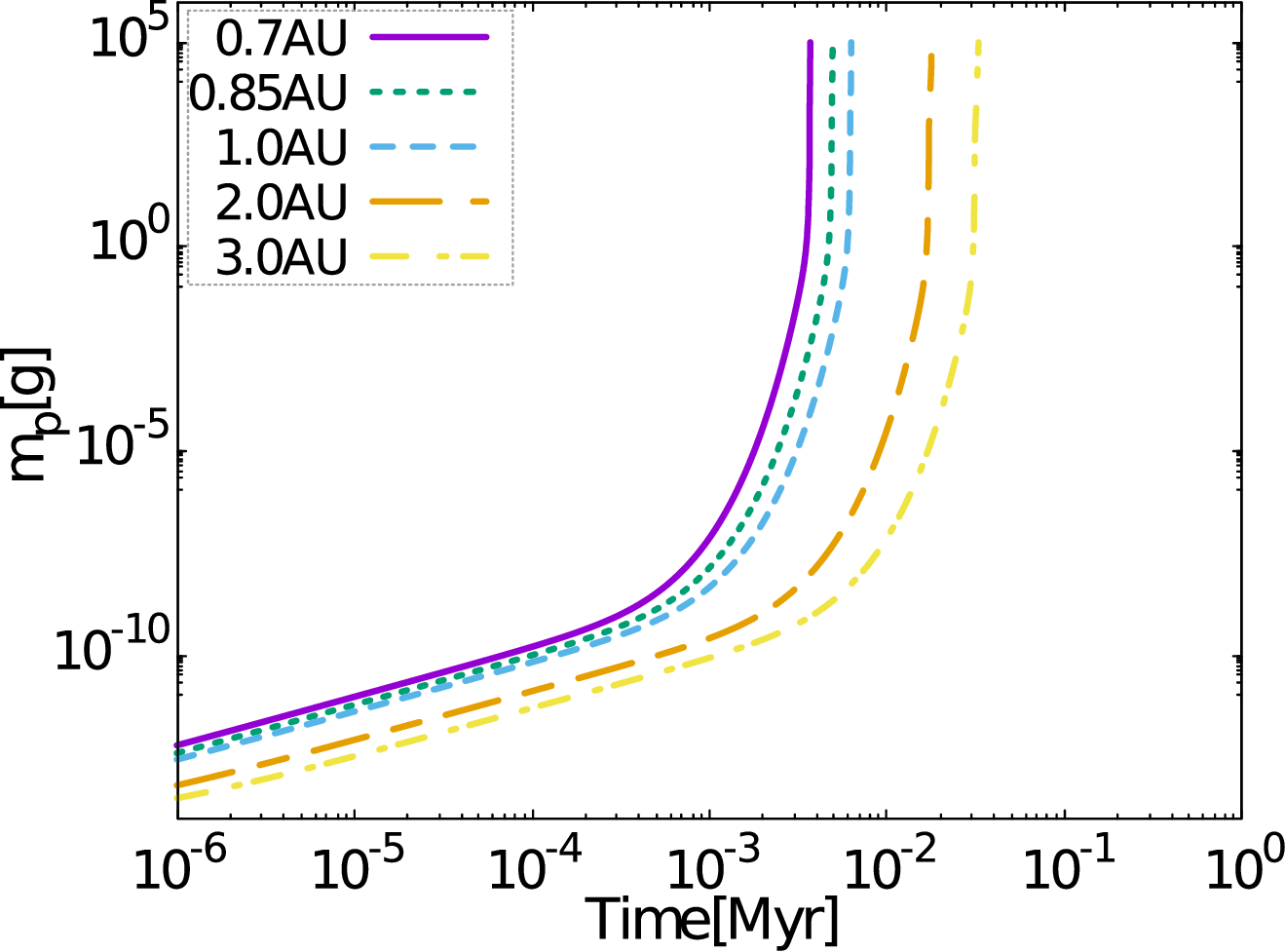}
 \caption{The time evolution of $m_{\mathrm{p}}$, starting from different initial distances from the central star. The accretion rate and the dust-to-gas fraction are $\dot{M}=10^{-7.0} ~M_{\odot}/\mathrm{yr}$ and $\bar{f}=\bar{f}_{\mathrm{MMSN}}$, respectively.} 
 \label{fig:dust_growth}
\end{figure}

Fig.\ref{fig:dust_growth} shows the time evolution of the mass of dust particles for $\dot{M}=10^{-7.0}~M_{\odot}/\mathrm{yr}$. Different curves in Fig.\ref{fig:dust_growth} show the growth of dust particles at different radii in the disk. For all cases, the initial dust radius is set to $0.1~\mu\mathrm{m}$. From Fig.\ref{fig:dust_growth}, we can calculate the timescale for a dust particle to grow to $m_{\mathrm{c}}$ at each radius $r$ within the disk. 
%Figure. 3
\par

As we described in section \ref{sec:Method:ParticleGrowth:model_assumption}, for simplicity, we assumed that dust particles grow through mutual collisions in situ, and once they have grown to $m_{\mathrm{c}}$, they subsequently drift towards the Sun due to the gas drag. However, one might consider that these assumptions would be oversimplified for, at least, our models. We present the results of a growth model that accounts for dust particle migration, which show the validity of in situ growth of dust particles.

According to \citet{1977MNRAS.180...57W} and \citet{1986Icar...67..375N}, the radial drift velocity of dust particles is given by
\begin{equation}
v_{\mathrm{rp}}=\frac{2\mathrm{St}}{1+\mathrm{St}^2}\eta r\Omega. \label{eq.38}
\end{equation}
To consider dust particles migration as a part of the dust growth model, we need to integrate eq.(\ref{eq.38}) at the same time as integrating eq.(\ref{eq.17}) by using eq.(\ref{eq.18}) to eq.(\ref{eq.37}).\par
%Figure. 3-3
\begin{figure}[hbtp]
 \includegraphics[width= 8cm]{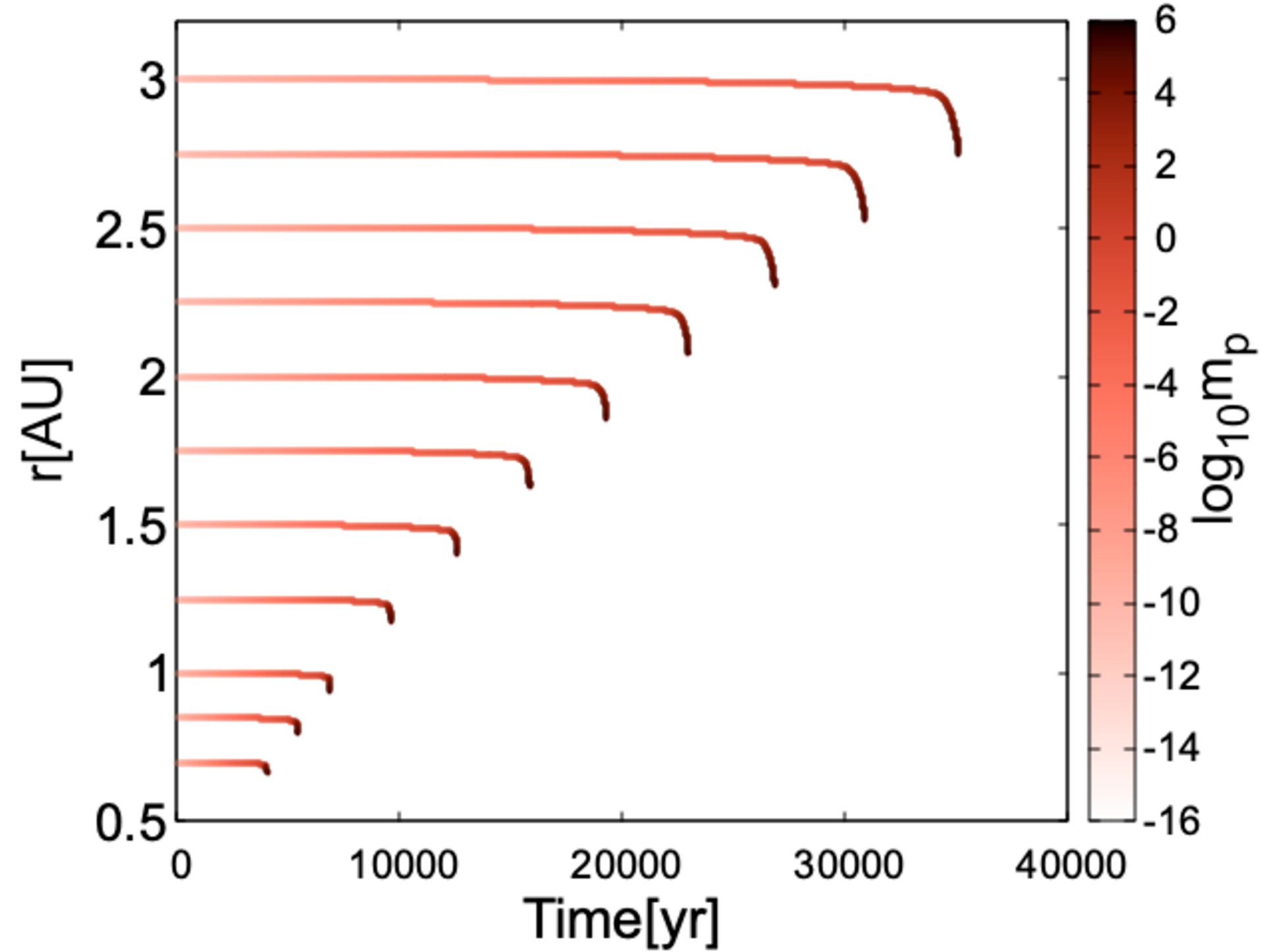}
 \caption{The time evolution of the particle distance from the central star $r_{\mathrm{p}}$, starting from different initial distances from the central star. The accretion rate and the dust-to-gas fraction are $\dot{M}=10^{-7.0}~M_{\odot}/\mathrm{yr}$ and $\bar{f}=\bar{f}_{\mathrm{MMSN}}$, respectively. The color of the line indicates dust particle mass (color bar). Different curves in Fig.\ref{fig:dust_migration} show the particles starting from different radii in the disk ranging from 0.7 to 3 AU.} 
 \label{fig:dust_migration}
\end{figure}
%Figure. 3-4
\begin{figure}[hbtp]
 \includegraphics[width= 8cm]{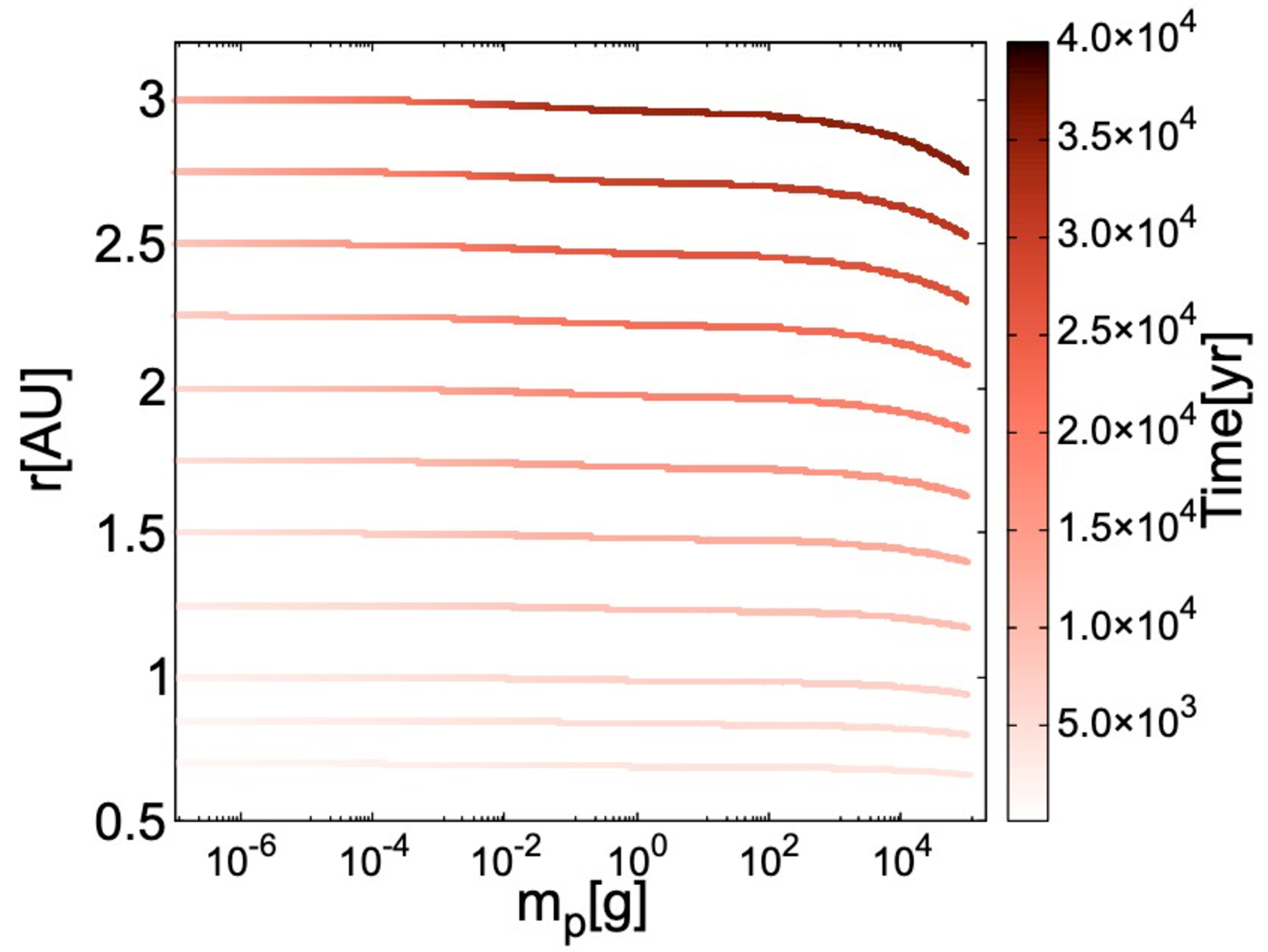}
 \caption{Change in dust particle distance from central star $r_{\mathrm{p}}$  with an increase in dust particle mass, starting from different initial distances from the central star. The color of the line indicates time (color bar).} 
 \label{fig:dust_migration_mass}
\end{figure}
Figs. \ref{fig:dust_migration} and \ref{fig:dust_migration_mass} show the evolution of the dust particle distance from the central star $r_{\mathrm{p}}$ as a function of time and mass of the dust particle, respectively. In Fig.\ref{fig:dust_migration}, the line color represents the logarithmic mass of the dust particles.
From these figures, it is seen that the radial migration of dust particles becomes pronounced when the dust particle mass exceeds $\sim10^2~{\rm g}$, and the mass growth becomes much faster than radial migration.

\subsubsection{The growth and migration timescales of dust particles}\label{sec:Method:ParticleGrowth:dust}
In section \ref{sec:Method:ParticleGrowth:mass}, we have seen that the evolution of the dust can be approximated as the two-stage phenomena. In the first stage, the dust grows at its radial position, until its mass reaches $m_{\mathrm{c}}$. In the second stage, it migrates inward.\par
Using eqs.(\ref{eq.17}) and (\ref{eq.38}), we can write the particle mass growth timescale  $t_{\mathrm{growth}}$ and the particle drift timescale
 $t_{\mathrm{drift}}$ as:
\begin{eqnarray}
t_{\mathrm{growth}}&=&\frac{m_{\mathrm{p}}}{\frac{dm_{\mathrm{p}}}{dt}}=\frac{m_{\mathrm{p}}}{\pi a_{\mathrm{p}}^2\rho_{\mathrm{p}}v_{\mathrm{rel,pp}}\left(1+\frac{2Gm_{\mathrm{p}}}{a_{\mathrm{p}}v_{\mathrm{rel,pp}}^2}\right)}, \label{eq.39} \\
t_{\mathrm{drift}}&=&\frac{(r_{\mathrm{p}}-r_{\mathrm{in}})}{v_{\mathrm{rp}}}=(r_{\mathrm{p}}-r_{\mathrm{in}})\frac{1+\mathrm{St}^2}{2(\mathrm{St})\eta r \Omega},  \label{eq.40}
\end{eqnarray}
where $r_{\mathrm{p}}$ is the particle distance from the central star and $r_{\mathrm{in}}$ is the location of IDB given by eq.(\ref{eq.11}).\par
\begin{figure}[hbtp]
 \includegraphics[width= 8.0cm]{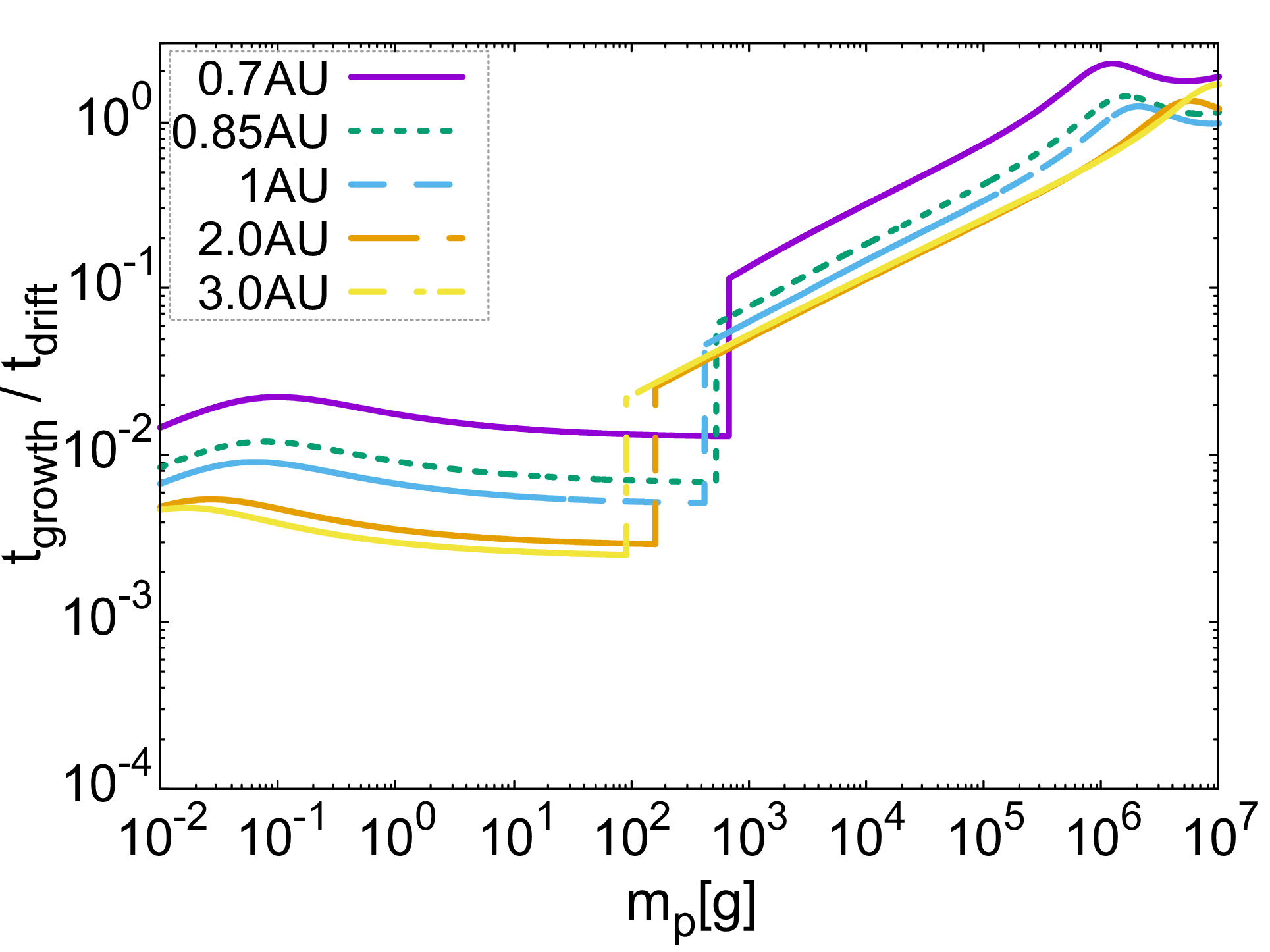}
 \caption{The ratio of particle growth time scale to particle drift timescale at different radii in the disk. The accretion rate and the dust-to-gas fraction are $\dot{M}=10^{-7.0} ~M_{\odot}/\mathrm{yr}$ and $\bar{f}=\bar{f}_{\mathrm{MMSN}}$, respectively. } 
 \label{fig:dust_time}
\end{figure}

Fig.{\ref{fig:dust_time}} shows the ratio of $t_{\mathrm{growth}}$ to $t_{\mathrm{drift}}$ as a function of the dust particle mass $m_{\mathrm{p}}$.
% %Figure. 4
The drift timescales are always longer than the mass growth timescale, as far as the dust mass is less than $\sim10^5$ g, regardless of the distance from the Sun.
Therefore, in this paper, we assume $m_{\mathrm{c}}=10^{5}$ g, and regards dusts with masses lager than $m_{\mathrm{c}}$ as pebbles.\par
\begin{figure}[hbtp]
 \includegraphics[width= 8.0cm]{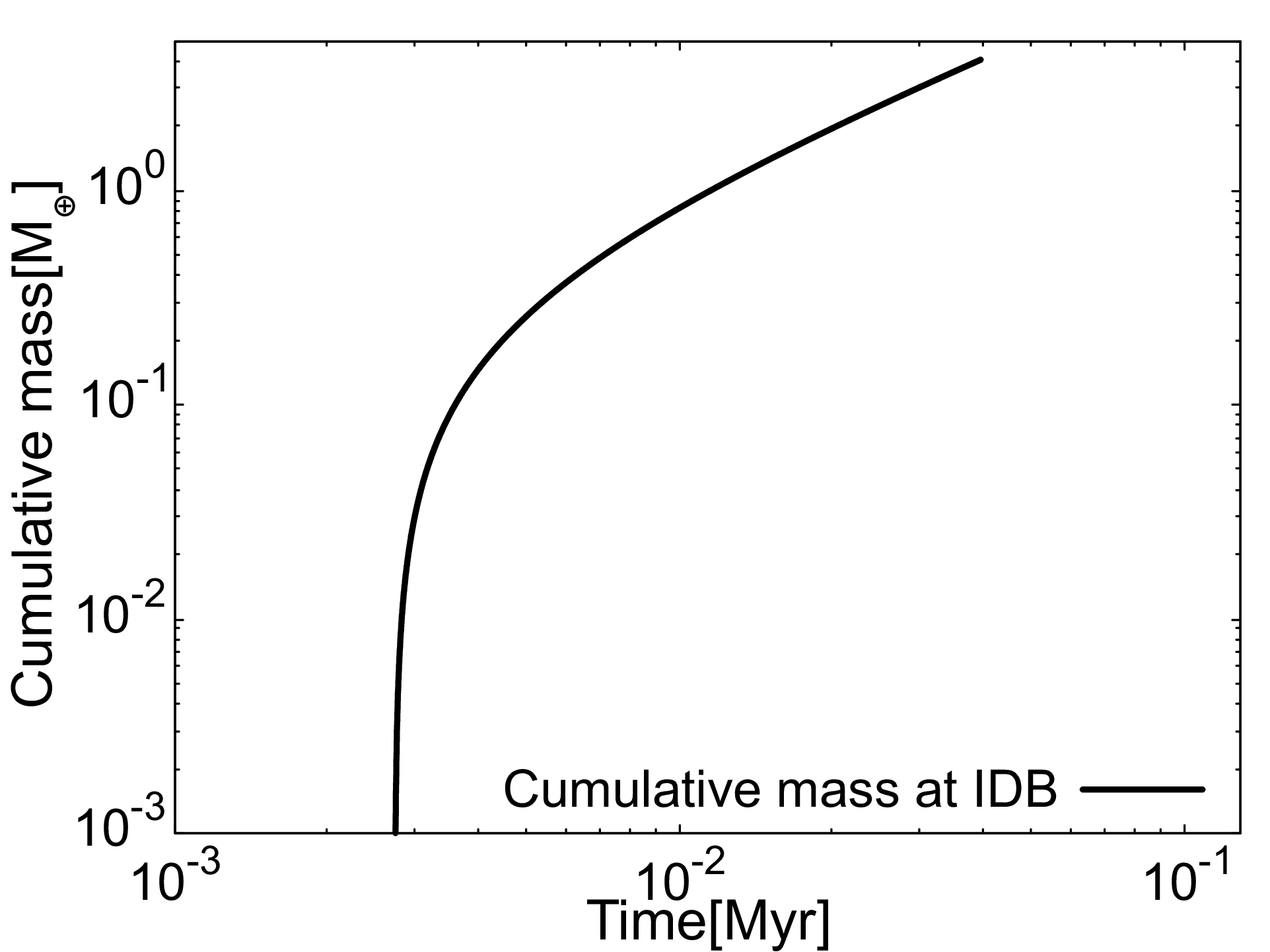}
 \caption{The cumulative mass reached IDB as a function of time. The accretion rate and the dust-to-gas fraction are $\dot{M}=10^{-7.0} ~M_{\odot}/\mathrm{yr}$ and $\bar{f}=\bar{f}_{\mathrm{MMSN}}$, respectively. } 
 \label{fig:mass_flux_theoretical}
\end{figure}

Fig.\ref{fig:mass_flux_theoretical} shows the cumulative mass reached to IDB as a function of time. In our model, we assume that all mass within WSZ eventually reaches IDB. Thus mass flow to IDB continues to time $t=3.7\times10^{4}$ yrs, with the inflow rate described as a function of time (see Appendix. \ref{sec:Appendix_mass_flux} for more details).
% %Figure. 4-2

\subsection{Numerical methods} \label{sec:Method:numerical}
Using the model for the gas disk and that for the growth and migration of dust particles discussed in sections \ref{sec:Method:gasdisk} and \ref{sec:Method:ParticleGrowth}, we can determine the dust accretion timescale onto IDB. The accretion of dust onto IDB leads to the formation of planetesimals and planets in the vicinity of IDB. This phase of planet formation involves non-linear evolution, in which gravity becomes important. Therefore, we conduct $N$-body simulations from this phase.\par
In this subsection, we first provide a brief description of the $N$-body code GPLUM \citep{2021PASJ...73..660I} which we used in this study. Then, we describe how to incorporate the disk model and the dust growth model discussed in sections \ref{sec:Method:gasdisk} and \ref{sec:Method:ParticleGrowth} into GPLUM. We describe how to set up the initial conditions for our $N$-body simulations.

\subsubsection{$N$-body simulation code GPLUM} \label{sec:Method:numerical:GPLUM}
GPLUM \citep{2021PASJ...73..660I} is a parallel $N$-body simulation code for studying the formation of planetary systems. This code is developed using Framework for Developing Particle Simulator (FDPS) 
\citep{2016PASJ...68...54I,2018PASJ...70...70N}, a general-purpose, high-performance library for particle simulations. GPLUM is highly scalable under a parallel environment. Hereafter, we briefly describe the numerical scheme used in GPLUM.\par

GPLUM uses the particle-particle particle-tree (P$^3$T) scheme, which is a hybrid integrator based on the splitting of the Hamiltonian \citep{2011PASJ...63..881O}. The Hamiltonian of the system is divided into two parts depending on the cutoff radii of particles. The gravitational force between two particles is divided into short-range and long-range terms using the switching function scaled by the cutoff radius. The short-range part is called the hard part, while the long-range is called the soft part. The hard part consists of gravitational potential due to the central star and short-range interactions of particles. The time integration of the hard part is integrated by the fourth-order Hermite scheme \citep{1991ApJ...369..200M} with the individual time-step scheme \citep{1963MNRAS.126..223A}. When there are no neighbor particles within the cutoff radius, an analytic solution of the Kepler orbit around the central star is used.
The soft part is calculated by using the Barnes-Hut tree scheme \citep{1986Natur.324..446B} available in FDPS. The hard and soft parts of the Hamiltonian used in the P$^3$T scheme are given by
\begin{eqnarray}
H&=&H_{\mathrm{Hard}}+H_{\mathrm{Soft}}, \label{Hamiltonian}\\
H_{\mathrm{Hard}}&=&\sum_{i}\left\lbrack\frac{|\boldsymbol{p}_i|^2}{2m_i}-\frac{GM_*m_i}{r_i}\right\rbrack\nonumber\\
&&-\sum_{i}\sum_{j>i}\frac{GM_*m_i}{r_{ij}}[1-W(r_{ij};r_{\mathrm{out}})], \label{Hard_part}\\
H_{\mathrm{Soft}}&=&-\sum_{i}\sum_{j>i}\frac{GM_*m_i}{r_{ij}}W(r_{ij};r_{\mathrm{out}}), \label{Soft_part}\\
r_{ij}&=&|\boldsymbol{r_{i}}-\boldsymbol{r}_j|, 
\end{eqnarray}
where $m_{i}$,$~\boldsymbol{p}_{i}$, $\boldsymbol{r}_{i}$, $r_{\mathrm{out}}$ and $W(r_{ij};r_{\mathrm{out}})$ are the mass, momentum, position of the $i$-th particle, the cutoff radius and the cutoff function for the Hamiltonian.

The cutoff radius for gravitational interactions between the $i$-th and $j$-th particles is set to be
\begin{eqnarray}
r_{\mathrm{out},ij}&=&\max(\tilde{R}_{\mathrm{cut},0}r_{\mathrm{Hill},i},\tilde{R}_{\mathrm{cut},1}v_{\mathrm{ran},i}\Delta t,
\tilde{R}_{\mathrm{cut},0}r_{\mathrm{Hill},j},\nonumber\\&&\tilde{R}_{\mathrm{cut},1}v_{\mathrm{ran},j}\Delta t), \label{eq.44}
\end{eqnarray}
where $\tilde{R}_{\mathrm{cut},0}$ and $\tilde{R}_{\mathrm{cut},1}$ are the parameters, $r_{\mathrm{Hill},i}$ and $r_{\mathrm{Hill},j}$ are the Hill radius, and $v_{\mathrm{ran},i}$ and $v_{\mathrm{ran},j}$ are the r.m.s random velocity for particles around $i$-th and $j$-th particles in GPLUM. Here, the random velocity is defined as the difference between its velocity and the Kepler velocity.

GPLUM uses the same cutoff function as  \citet{2017PASJ...69...81I} which is defined by
\begin{equation}
W(y;\gamma)=\left\{
\begin{array}{ll}
\frac{7(\gamma^6-9\gamma^5+45\gamma^4-60\gamma^3\ln\gamma-45\gamma^2+9\gamma-1)}{3(\gamma-1)^7}y, &\\
    \hspace{4.0cm} (y<\gamma),&\\
    f(y;\gamma)+[1-f(1;\gamma)y], &\\
    \hspace{3.5cm}(\gamma\leq y<1),&\\
        1, &\\
       \hspace{4.0cm} (1\leq y), &\\
\end{array}\right.
\end{equation}
where
\begin{eqnarray*}
f(y;\gamma)=&\lbrace-10/3y^7+14(\gamma+1)y^6-21(\gamma^2+3\gamma+1)y^5\\
            &+\lbrack 35(\gamma^3+9\gamma^2+9\gamma+1)/3\rbrack y^4\\
            &-70(\gamma^3+3\gamma^2+\gamma)y^3\\
            &+210(\gamma^3+\gamma^2)y^2-140\gamma^3y\ln y\\
            &+(\gamma^7-7\gamma^6+21\gamma^5-35\gamma^4)\rbrace/(\gamma-1)^7.
\end{eqnarray*}
The cutoff function becomes unity when $r_{ij}$ is longer than the cutoff radius for gravitational interactions between the $i$-th and $j$-th particles ($r_{ij}>r_{\mathrm{out},ij}$). Therefore, gravitational interactions of the hard part work only between particles with $r_{ij}<r_{\mathrm{out,ij}}$.\par

Using the individual cutoff method, GPLUM has made it possible to split gravitational interactions more efficiently compared to the original scheme that uses the ``shared cutoff" radius for all particles. GPLUM can handle a large number of particles ($N\sim10^6$), a wide range in masses of particles, and a wide radial range (more than several AU) in simulations since it adopts the P$^3$T scheme with the individual cutoff radius method.

\subsubsection{The prescription for our $N$-body simulations} \label{sec:Method:numerical:prescription}
We aim to study the planet formation process in the vicinity of IDB, taking into account the in-situ growth of dust and dust migration discussed in section \ref{sec:Method:ParticleGrowth}. Thus, in this section, we describe how to implement (1) the in-situ growth of dust, (2) the dust migration due to gas drag, and (3) the planet formation in the vicinity of IDB in GPLUM.

In our $N$-body simulations, we initially set particles with masses ranging from $\sim 10^{22}$ g to $\sim10^{24}$ g, depending on the number of particles used in each simulation to match the solid surface density distribution of the disk (see section \ref{sec:Method:numerical:initial} and Appendix \ref{sec:Appendix_mass_flux}). Each of these particles represents a swarm of pebbles with a mass of $m_{\mathrm{c}}$. We reproduce the in-situ growth of dust by maintaining the placed particles in-situ until they reach the timescale for the dust to grow up to $m_{\mathrm{c}}$. To realize the in-situ orbital motion of particles, we need to suppress the effects of gravitational interaction between particles. Therefore, we set the mass of each particle to $10^{14}$ g, which is the minimum mass that can be handled in GPLUM to minimize the effects of inter-particle gravitational interactions. Furthermore, the gas drag is disabled for each particle until it reaches the timescale for the dust mass to reach $m_{\mathrm{c}}$. 

 Once the growth timescale is reached at each position within the disk, the gas drag force is applied to each particle to match the dust accretion timescale to reproduce mass accretion onto IDB (see Appendix \ref{sec:Appendix_mass_flux}). The gas drag force $F_{\mathrm{drag}}$ is given by \citet{1976PThPh..56.1756A} and \citet{PhysRev.23.710}:
\begin{equation}
F_{\mathrm{drag}}=  \left\{
			\begin{array}{ll}
				-\frac{4\pi}{3}\sqrt{\frac{8}{\pi}}\rho_{\mathrm{g}}a_{\mathrm{p}}^2c_{\mathrm{s}}v_{\mathrm{rel,pp}} &\\\mathrm{For~Epstein~region} ~(a_{\mathrm{p}}<\frac{9}{4}\lambda) \\
				-6\pi\rho_{\mathrm{g}}\nu a_{\mathrm{p}}v_{\mathrm{rel,pp}} &\\ \mathrm{For~Stokes~region} ~(a_{\mathrm{p}}>\frac{9}{4}\lambda).
			\end{array}
		\right. \label{eq.46}
\end{equation}
Note that $a_{\mathrm{p}}$ in eq.(\ref{eq.46}) is the radius of a particle with mass $m_{\mathrm{c}}$, since we are considering timescales for the accretion of pebble-sized objects within our simulations. 

When a particle passes through IDB, the particle's radial migration is stopped by cutting off the gas drag force $F_{\mathrm{drag}}$, and the mass is returned to the actual mass given in the initial condition in order to reproduce pebble accretion and accumulation onto IDB. This is equivalent to treating particles passing through IDB as having grown to planetesimals with the mass ($\sim10^{22-24}$ g) initially given to them due to the gravitational instability caused by dust concentration in the vicinity of IDB (e.g., \cite{2010ApJ...714.1155K}; \cite{2014ApJ...780...53C}). In the same manner, when particles undergoing radial migration merge with particles that have already passed through IDB, they are processed to return to the mass given in the initial condition before merging particles. Here we note that in our $N$-body simulations, we assume perfect accretion during particle collisions, neglecting the effects of fragmentation.

\subsubsection{Initial condition settings}\label{sec:Method:numerical:initial}
In our simulations, we placed $6\times10^4 \sim 1\times 10^6$ particles of equal mass $m_{\mathrm{p}}$ in the radial range of $r_{\mathrm{in}}=0.58$ AU to $r_{\mathrm{WSZ}}=3.28$ AU. Below, we provide a detailed description of the initial configuration of the dust particles.

The total dust mass in the radial range of $r_{\mathrm{in}}$ to $r_{\mathrm{WSZ}}$ is given by
\begin{equation}
M_{\mathrm{d,tot}}=\int_{r_{\mathrm{in}}}^{r_{\mathrm{WSZ}}}2\pi r\Sigma_{\mathrm{p}} dr. \label{total_dust}
\end{equation}
Using eq.(\ref{total_dust}), $m_{\mathrm{p}}$ is given by
\begin{equation}
m_{\mathrm{p}}=\frac{M_{\mathrm{d,tot}}}{N}.  \label{particle_mass}
\end{equation}
The radial position of $i$-th particle, $r_{i}$ is given by
\begin{equation}
M(r)=\int_{r_{i-1}}^{r_{i}}2\pi r\Sigma_{\mathrm{p}} dr=m_{\mathrm{p}}, \label{eq.45}
\end{equation}
where $r_{1}=r_{\mathrm{in}}$.

Table.\ref{tab1} summarizes our initial models. The surfaces of Table.\ref{tab1} show the names of the models, the dust-to-gas fraction $\bar{f_{\mathrm{p}}}$, the initial total dust mass $M_{\mathrm{tot}}$, the initial number of particles $n_{\mathrm{p}}$, the initial mass of a particle $m_{\mathrm{p,init}}$, and the mass of the largest particle at the end of the calculation $m_{\mathrm{p,max}}$. Our reference model is N120Kc1, with $N=120$k. For this model, the dust fraction normalized by MMSN value, $f$, is unity. We made three series of models starting from this reference model. In the first series, we vary $f$ from 0.25 to 4. (N120Ka to Ke). In the second series, we vary the number of particles from 60k to 1M. In the third series, we added two models with all parameters the same as the reference model, but with different initial random seeds for angular coordinates of particles.

\begin{table*}
  \tbl{List of models.}{%
  \begin{tabular}{cccccc}
      \hline
      Name & $\bar{f}(\bar{f}_{\mathrm{MMSN}})$ & $M_{\mathrm{tot}} (M_{\oplus})$& \ $N_{\mathrm{p}}$ & $m_{\mathrm{p,init}} (M_{\oplus})$ & $m_{\mathrm{p,max}} (M_{\oplus})$  \\ 
      \hline
       N120Kc1& $1$ & $4.07$ & $1.2\times10^5$& $3.3\times10^{-5}$ & $2.4$ \\
       N120Ka& $0.25$ & $1.02$ & $1.2\times10^5$& $8.2\times10^{-6}$ & $0.49$ \\
       N120Kb& $0.5$ & $2.03$ & $1.2\times10^5$& $4.1\times10^{-6}$ & $1.4$ \\
       N120Kd& $2$ & $8.14$ & $1.2\times10^5$& $6.6\times10^{-5}$ & $5.4$ \\
       N120Ke& $4$ & $16.27$ & $1.2\times10^5$& $1.3\times10^{-4}$ & $7.2$ \\
       N60Kc& $1$ & $4.07$ & $6.0\times10^4$& $6.6\times10^{-5}$ & $2.6$ \\
       N200Kc& $1$ & $4.07$ & $2.0\times10^5$& $2.0\times10^{-5}$ & $2.1$ \\
       N1Mc& $1$ & $4.07$ & $1.0\times10^6$& $4.0\times10^{-6}$ & $2.5$ \\
       N120Kc2& $1$ & $4.07$ & $1.2\times10^5$& $3.3\times10^{-5}$ & $2.7$ \\
       N120Kc3& $1$ & $4.07$ & $1.2\times10^5$& $3.3\times10^{-5}$ & $2.5$ \\
      \hline
    \end{tabular}}\label{tab1}
\end{table*}

\section{Results} \label{sec:Results}
In section \ref{sec:Results:formation}, we summarize how planet formation proceeds in our model by showing the distribution of planets, the time evolution of mean and maximum masses of planets, and the cumulative mass distribution, using our fiducial model N120Kc1 as an example. 
In section \ref{sec:Results:evolution}, we compare simulation results for disks with different dust-to-gas fractions. 
In section \ref{sec:Results:number}, we investigate the particle number dependence of our simulations by using $\bar{f}=\bar{f}_{\mathrm{MMSN}}$ disk.
 %Figure. 8
\begin{figure*}[hbtp]
 \includegraphics[width= 17cm]{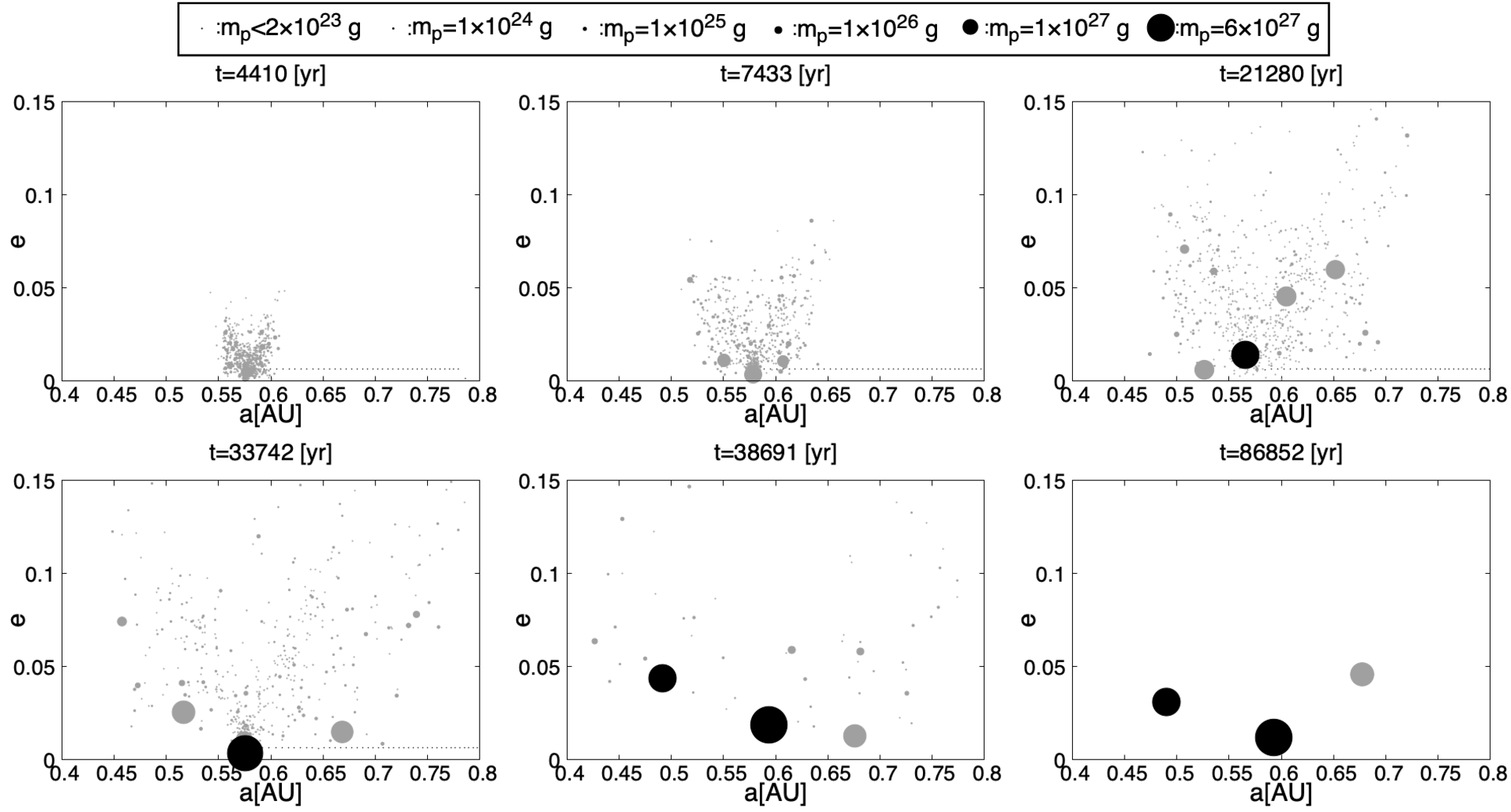}
 \caption{Snapshots of planetary formations in model N120Kc1 arranged in chronological order from the upper left panel to the lower right panel in the $a-e$ plane. Bodies above Earth mass are filled in black. The boxed legend shows the size of each representative mass and the filled circle size is proportional to $m_{\mathrm{p}}^{1/3}$. } 
 \label{fig:planet_formation}
\end{figure*}

\subsection{Planetary formation in our model}\label{sec:Results:formation}
In this subsection, we first present the results of model N120Kc1 and give an overview of the planet formation process in our model. 

\subsubsection{Planet formation process in Model N120Kc1} \label{sec:Results:formation:model}
Fig.\ref{fig:planet_formation} shows the distribution of particles in
the $a$-$e$ plane at several different epochs, where $a$ is the semi-major
axis and $e$ is the eccentricity. The sizes of points are proportional
to $m_{\mathrm{p}}^{1/3}$, where $m_{\mathrm{p}}$ is the mass of a particle.

We can see that massive protoplanet forms very early on ($t=4410$ yrs). In
this run, one massive planet continues to grow, but several less
massive planets are formed and grow. At the end of the simulation,
three planets remained.

\begin{figure}[hbtp]
 \includegraphics[width= 8cm]{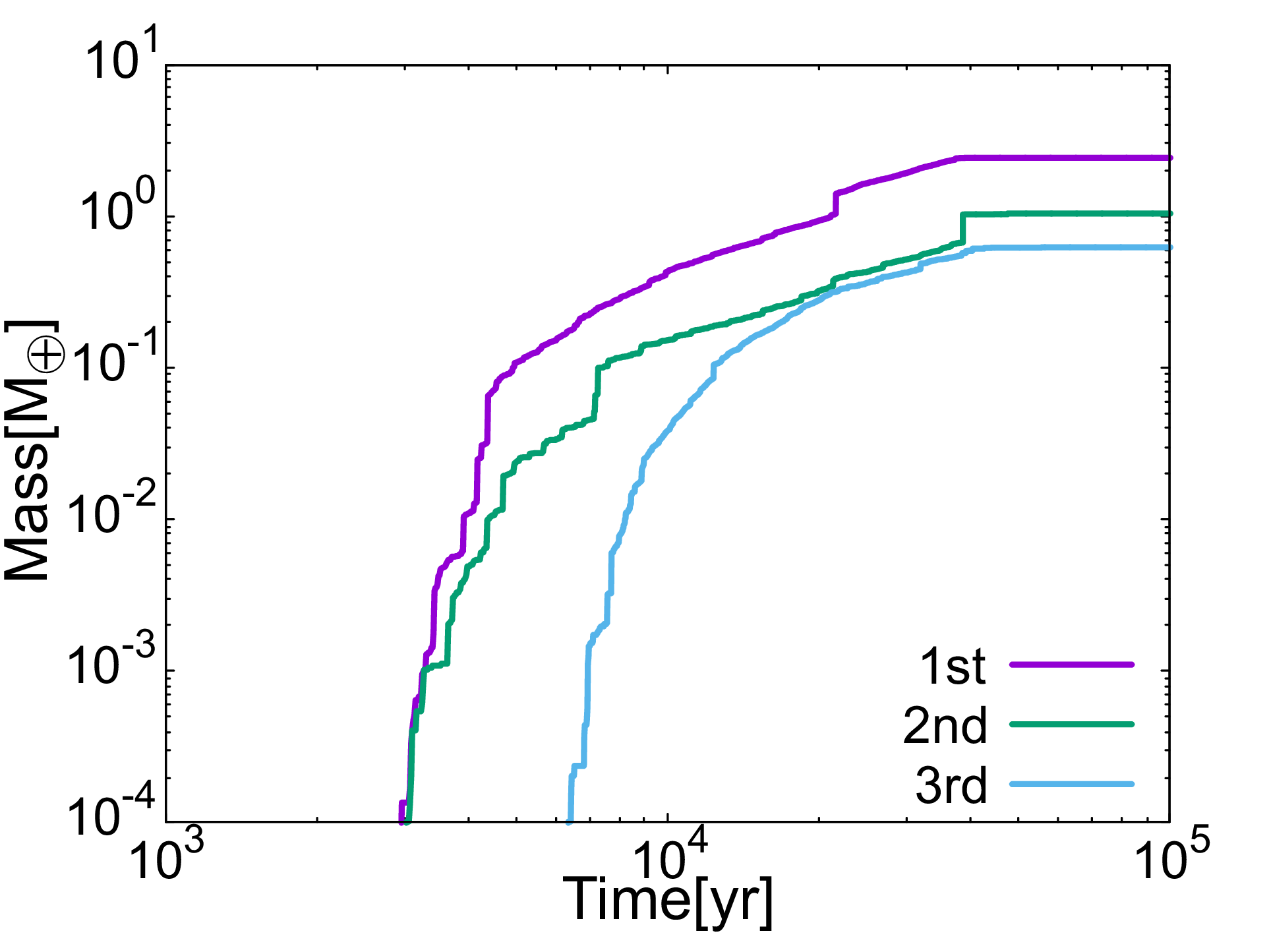}
 \caption{The mass evolution of the three largest planets as a function of time in model N120Kc1.} 
 \label{fig:mass_evolution}
\end{figure}

Fig.\ref{fig:mass_evolution} shows the evolution of the masses of the three most massive
particles. The mass of the most massive planet jumps at $t=
2.16\times10^4$ yrs. Here, planets with masses $1.0~M_{\oplus}$ and $0.36~M_{\oplus}$
collided.  A similar jump also occurs at $t=3.9\times10^4$ yrs. Here, planets with
masses $0.66~M_{\oplus}$ and $0.35 M_{\oplus}$ collided. These collisions
can be regarded as Giant impacts, which might result in the formation
of large moons.

\begin{figure*}[htbp]
 \centering
 \includegraphics[width= 17cm]{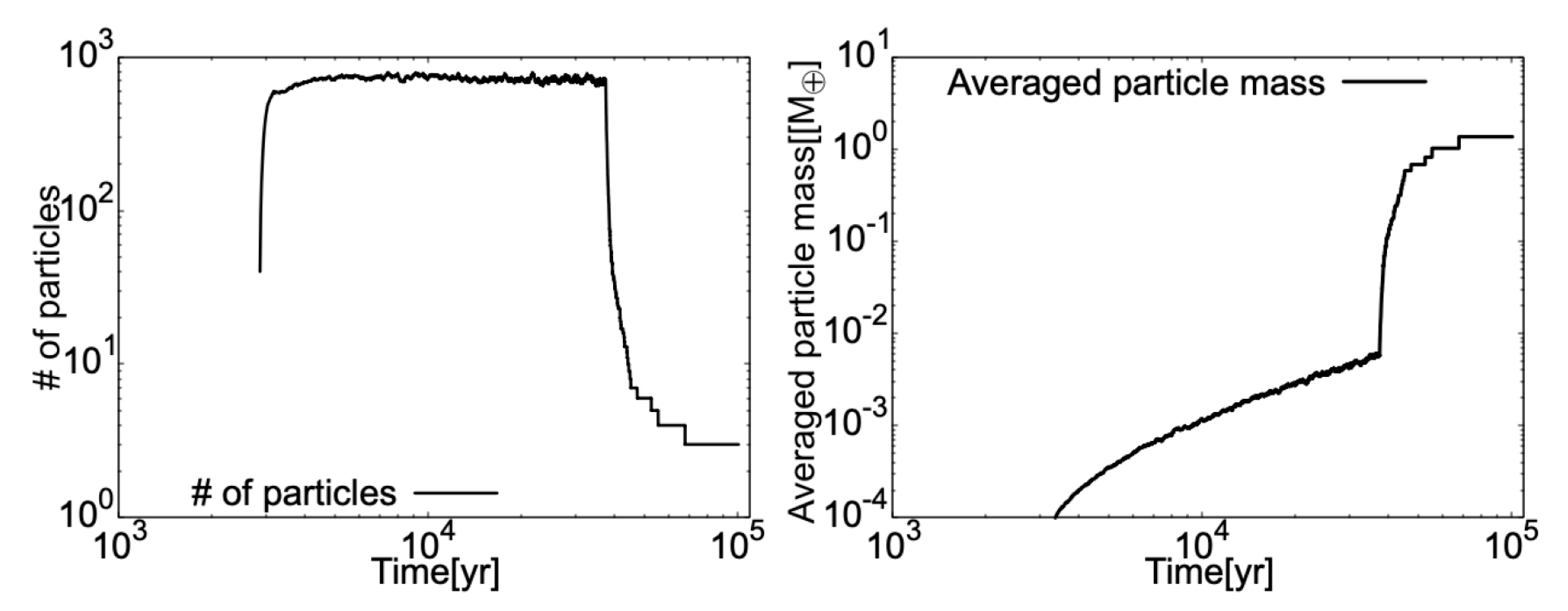}
 \caption{The number of particles that are above the planetesimal mass (left) and 
 the average mass of particles (right) as a function of time.}
 \label{fig:number_average}
\end{figure*}

Fig.\ref{fig:number_average} shows the time evolution of the number of
planetesimals and the average particle mass. The inflow of pebbles ends
at $t=3.7\times10^4$ yrs. As we can see, while the pebble inflow continues, the
total number of particles remains almost  constant, and the average mass
of particles grows nearly linearly, After the inflow stopped,
the number of remaining particles decreases quickly and as a result the
average particle mass goes up quickly.

These apparent changes simply reflect the fact that the pebble inflow
stopped, and do not really imply the change in the growth mode. 
The late-phase evolution of the averaged particle mass is characterized by sudden increases, as a consequence of the giant impacts.
 %Figure. 9
\begin{table*}[hbtp]
  \tbl{Summary of models with different dust to gas ratios $\bar{f}$ .\footnotemark[$*$]}{%
  \begin{tabular}{cccccc}
      \hline
      Name& $\bar{f}(\bar{f}_{\mathrm{MMSN}})$ & $M_{1\mathrm{st}} (M_{\oplus})$& $M_{2\mathrm{nd}} (M_{\oplus})$& $M_{3\mathrm{rd}} (M_{\oplus})$ & $M_{4\mathrm{th}} (M_{\oplus})$\\ 
      \hline
       N120Ka& $0.25$ & $0.49$ & $0.25$& $0.23$& 0.03 \\
       N120Kb& $0.5$ & $1.4$ & $0.37$& $0.25$& - \\
       N120Kc1& $1$ & $2.4$ & $1.0$& $0.60$& - \\
       N120Kd& $2$ & $5.4$ & $1.4$& $1.4$& - \\
       N120Ke& $4$ & $7.2$ & $6.1$& $1.6$&  $1.34$  \\
      \hline
    \end{tabular}}\label{tab2}
\begin{tabnote}
\footnotemark[$*$] From left to right, the surfaces show the names of the models, the dust-to-gas fraction ($\bar{f_{\mathrm{p}}}$), the heaviest particle mass ($M_{1\mathrm{st}}$), the second heaviest particle mass ($M_{2\mathrm{nd}}$), the third heaviest particle mass ($M_{3\mathrm{rd}}$), the fourth heaviest particle mass ($M_{4\mathrm{rd}}$).\\
\end{tabnote}
\end{table*} 

\begin{table*}[hbtp]
  \tbl{Summary of models with different particle numbers $N_{\mathrm{p}}$.\footnotemark[$*$]}{%
  \begin{tabular}{ccccccc}
      \hline
      Name& $N_{\mathrm{p}}$ & $M_{1\mathrm{st}} (M_{\oplus})$& $M_{2\mathrm{nd}} (M_{\oplus})$& $M_{3\mathrm{rd}} (M_{\oplus})$ & $M_{4\mathrm{th}} (M_{\oplus})$ & $M_{5\mathrm{th}} (M_{\oplus})$\\ 
      \hline
       N60Kc& $6\times10^4$ & $2.67$ & $0.81$& $0.60$& - & - \\
       N120Kc1& $1.2\times10^5$ & $2.4$ & $1.0$& $0.60$& - & - \\
       N200Kc& $2\times10^5$ & $2.1$ & $0.72$& $0.62$& $0.54$ & $0.09$ \\
       N1Mc& $1\times10^6$ & $2.5$ & $0.92$& $0.62$& - & -  \\
      \hline
    \end{tabular}}\label{tab3}
\begin{tabnote}
\footnotemark[$*$] From left to right, the surfaces show the names of the models, the initial number of particles ($N_{\mathrm{p}}$), the heaviest particle mass ($M_{1\mathrm{st}}$), the second heaviest particle mass ($M_{2\mathrm{nd}}$), the third heaviest particle mass ($M_{3\mathrm{rd}}$), the fourth heaviest particle mass ($M_{4\mathrm{rd}}$), the fifth heaviest particle mass ($M_{5\mathrm{rd}}$).\\
\end{tabnote}
\end{table*} 
\subsubsection{Mass distribution of planetesimals and protoplanets}\label{sec:Results:formation:mode}
\begin{figure}[htbp]
 \centering
 \includegraphics[width= 8cm]{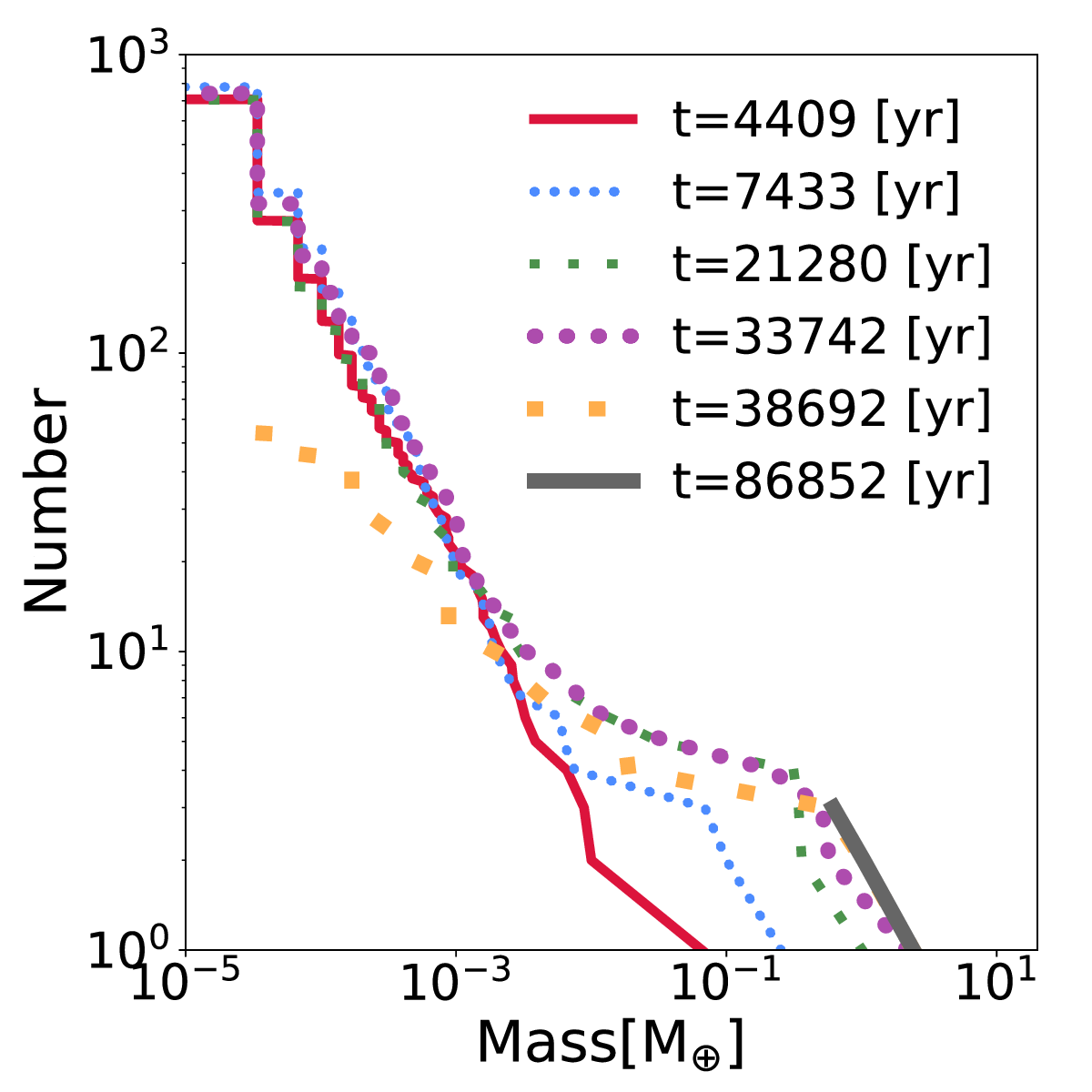}
 \caption{The time evolution of the cumulative mass distribution in model N120Kc1 (Note that particles that have not yet accreted to IDB are ignored)} 
 \label{fig:cumulative_mass}
\end{figure}

Fig.\ref{fig:cumulative_mass} shows the cumulative mass distribution of planetesimals for the same epochs as in Fig.\ref{fig:planet_formation}. We can see that the evolution of the mass distribution in our model is very similar to that in the classical model starting from MMSN ignoring infalls driven by aerodynamic drag or type-I migration (\authorcite{1996Icar..123..180K} \yearcite{1996Icar..123..180K,1998Icar..131..171K}, \cite{1998NewA....3..411M}). 
Initially, a power-law distribution develops,
and then a few massive bodies start to grow, leaving out other less massive bodies. After the pebble infall stopped, these few massive bodies eat up the remaining small bodies.

This is because pebbles are continuously supplied from MSZ to IDB  in our model. Thus, in our model, a few  massive planets are formed while the pebble accretion is taking place. As a result, the formation time of Earth-like planets is much shorter compared to the prediction of the classic model, where Earth-like planets are formed through collisions between protoplanets after the planetesimals are cleared out. 
 %Figure. 10
%
 %Figure. 11

\subsection{The effect of the  dust fraction}\label{sec:Results:evolution}
\begin{figure*}[htbp]
 \begin{center}
 \includegraphics[width= 17.0cm]{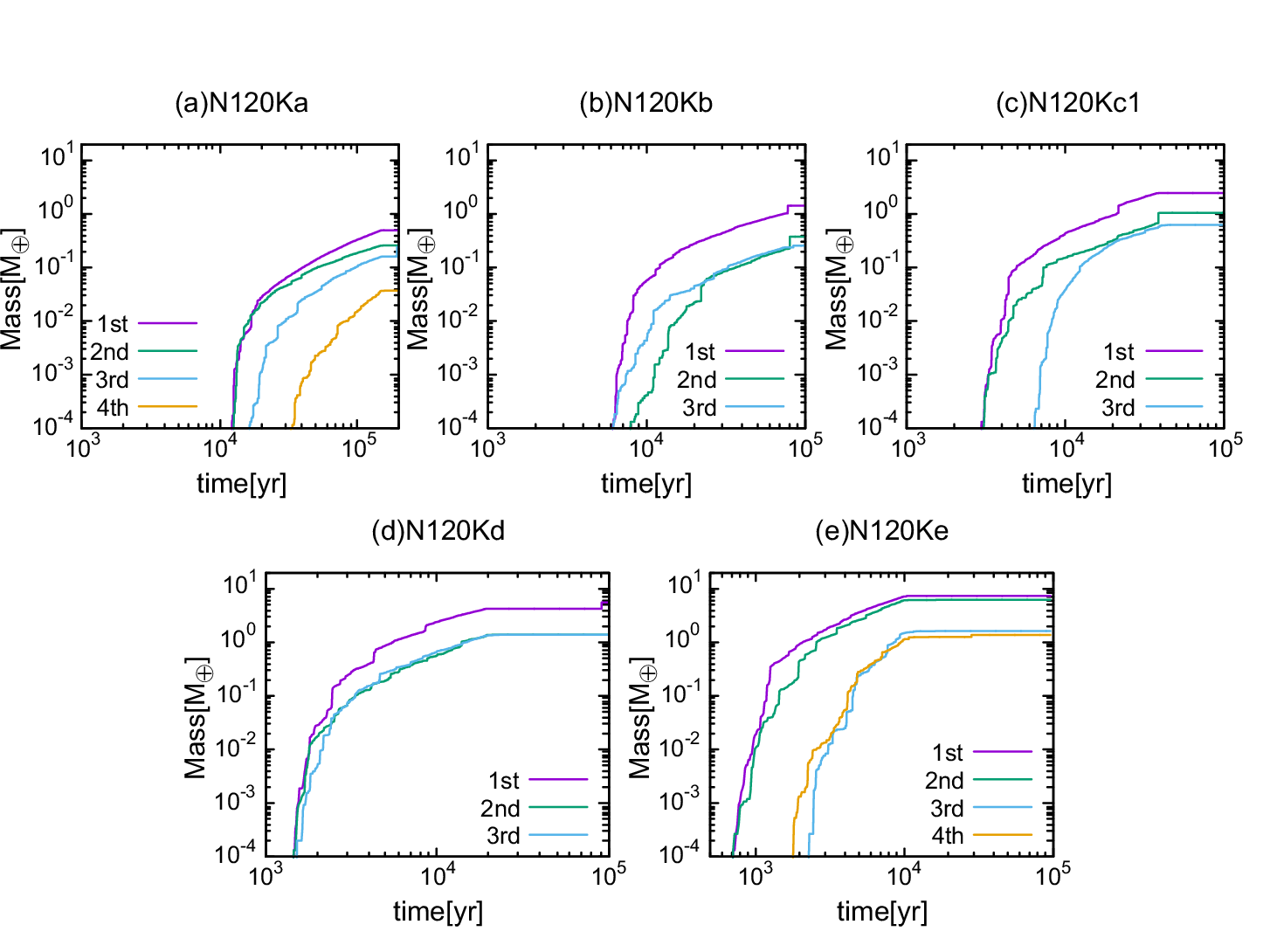}
 \end{center}
 \caption{Time evolution of particles' mass for each dust-to-gas fraction in the $m_{\mathrm{p}}-t$ plane. Results are shown for disks of $\bar{f}=(0.25, 0.5, 1, 2, 4)\times\bar{f}_{\mathrm{MMSN}}$ in order from upper left to the lower right in the figure, respectively.} 
 \label{fig:various_mass_evolution}
\end{figure*}

Fig.\ref{fig:various_mass_evolution} shows the evolution of the masses of three (or four) most massive bodies for runs with different initial dust fraction $\bar{f}$. In our model, the dust growth timescale depends on the dust number density and thus on $\bar{f}$, and is shorter for larger $\bar{f}$. Thus, the infall timescale is also shorter for larger $\bar{f}$. This is why the growth of massive bodies is faster for larger $\bar{f}$.

We can see that, even though the mass inflow rate and total mass are changed by a factor of 16, the final outcomes are surprisingly similar. In all cases, three (or four in one case) massive bodies are formed and their masses are within a factor of five. 

\begin{figure}[hbtp]
 %\centering
 \includegraphics[width=8cm]{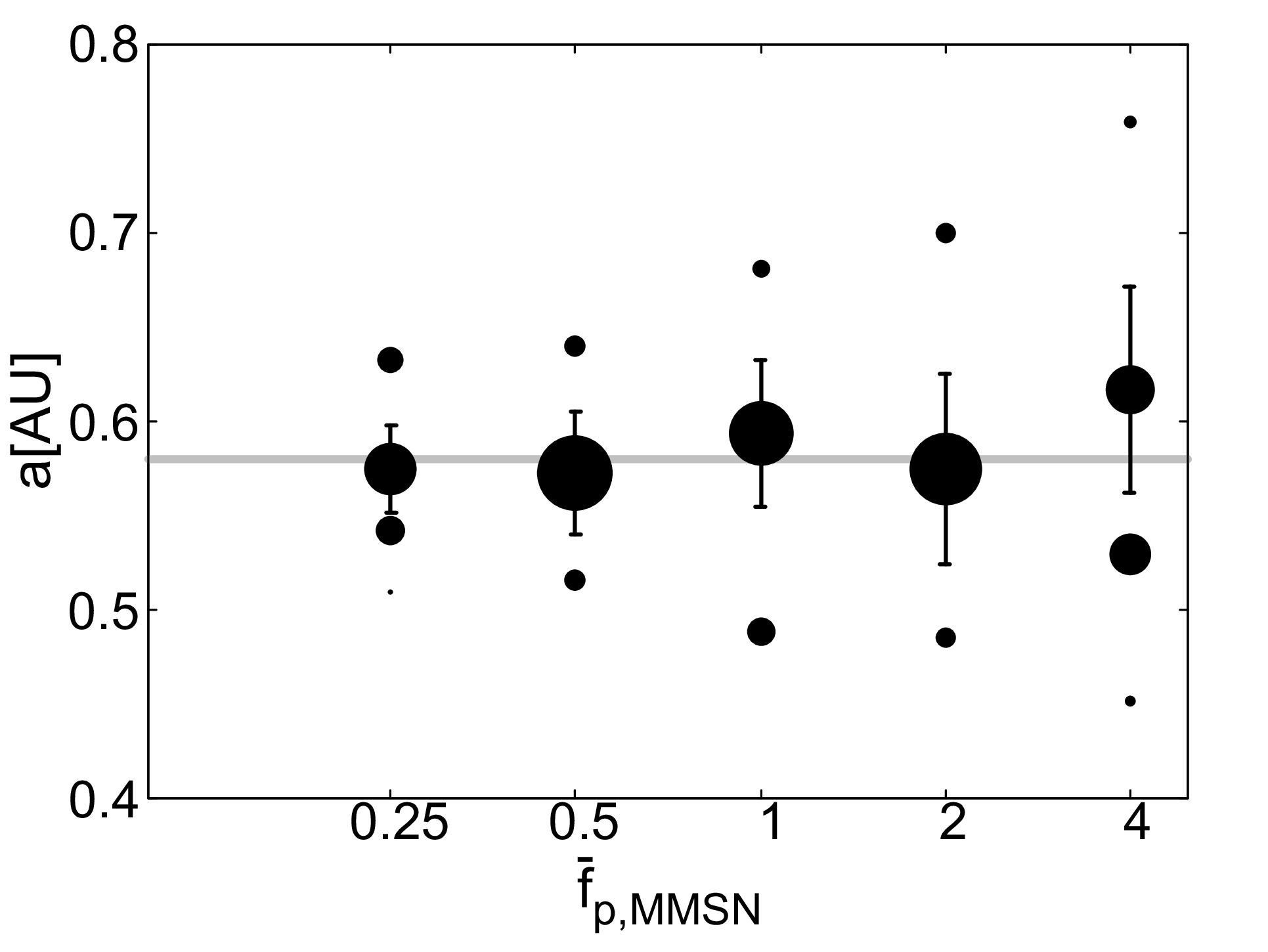}
 \caption{The semi-major axis and mass of massive bodies at the end of the simulation for models with different $\bar{f}$. The size of each dot represents the planet mass normalized by the total dust mass for each $\bar{f}$. The lines from the center of the most massive planets to both sides have the length of 5$r_{\mathrm{H}}$. The grey line represents inner dead zone boundary ($\sim 0.58$AU). See table \ref{tab2} for detailed values of the planetary masses formed at each value of $\bar{f}$. } 
 \label{fig:semi_major_comparison}
\end{figure}

Fig.\ref{fig:semi_major_comparison} shows the semi-major axis and mass of massive bodies at the end of the simulation for models with different $\bar{f}$. We show the theoretical curve of the planet separation,
\begin{equation}
  a = r_{\mathrm{p}} \pm 5 r_{\mathrm{H}}, \label{eq.Hill_radius}
\end{equation}
where $r_{\mathrm{H}}$ is the Hill radius of the most massive body.  We can see that the dependence of final separation to $\bar{f}$ is due to the difference in the Hill radius.

\begin{figure*}[htbp]
 \begin{center}
 \includegraphics[width= 15.5cm]{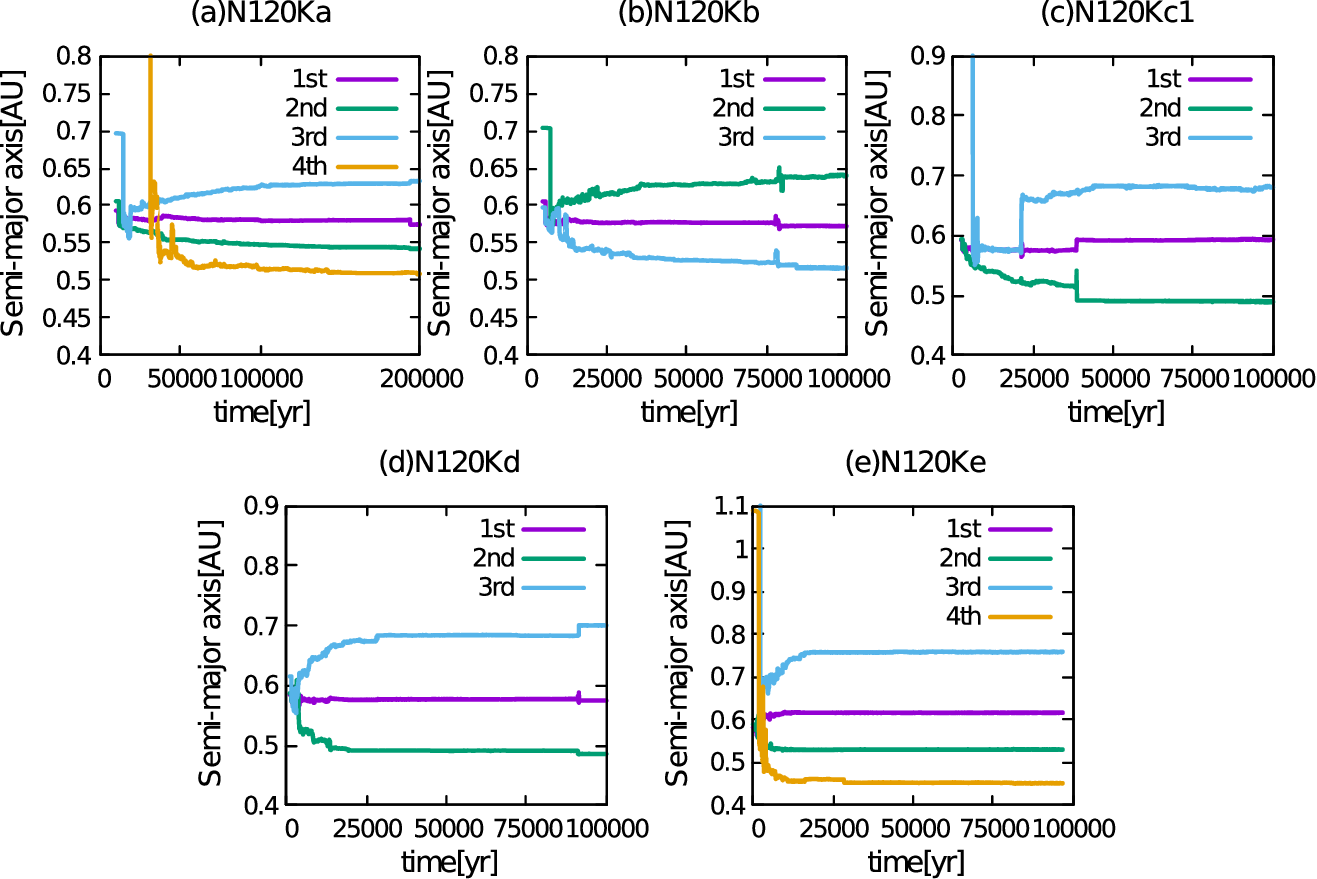}
 \end{center}
 \caption{Time evolution of particles' semi-major axis for each dust-to-gas fraction in the $a-t$ plane. Results are shown for disks of $\bar{f}=(0.25, 0.5, 1, 2, 4)\times\bar{f}_{\mathrm{MMSN}}$ in order from upper left to the lower right in the figure, respectively.} 
 \label{fig:various_semi_major}
\end{figure*}

Fig.\ref{fig:various_semi_major} shows the evolution of the semi-major axes of the massive bodies. We can see that the final separation of bodies is larger for larger $\bar{f}$. This is not surprising since the final mass of bodies is larger for $\bar{f}$.

\begin{figure*}[hbtp]
 \begin{center}
 \includegraphics[width= 15.5cm]{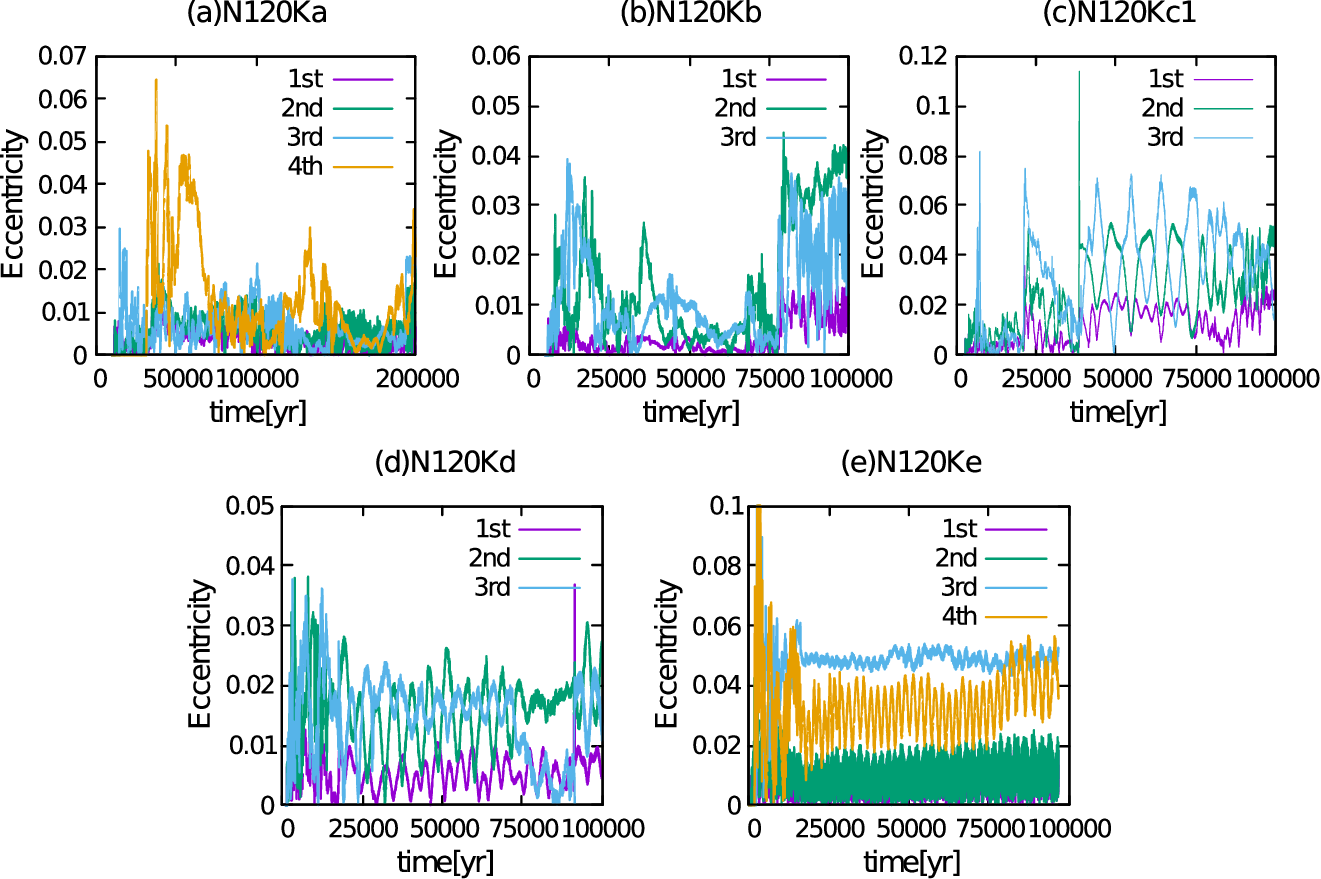}
 \end{center}
 \caption{Time evolution of particles' eccentricity for each dust-to-gas fraction in the $e-t$ plane. Results are shown for disks of $\bar{f}=(0.25, 0.5, 1, 2, 4)\times\bar{f}_{\mathrm{MMSN}}$ in order from upper left to the lower right in the figure, respectively. }
 \label{fig:various_eccentricity}
\end{figure*}

Fig.\ref{fig:various_eccentricity} shows the evolution of eccentricities of massive bodies. We can see that the eccentricities are generally small, even for the case of very massive planets. This is again quite different from the prediction of the classical model, which requires that the eccentricities of protoplanets grow large enough to allow mutual collisions. In our model, massive bodies are formed by pebble accretion, and there is no need  to let them collide. As we have seen in section \ref{sec:Results:formation:model}, there are a few collision events that can be regarded as giant impacts. However, even when such collisions take place, the eccentricities of bodies are not very large, simply because their radial separation is small.

%

%Figure. 17

%Figure. 13
%
%Figure. 14

\subsection{The effect of the mass resolution}\label{sec:Results:number}
 %Figure. 15
\begin{figure*}[hbtp]
 %\centering
 \begin{center}
 \includegraphics[width= 16.0cm]{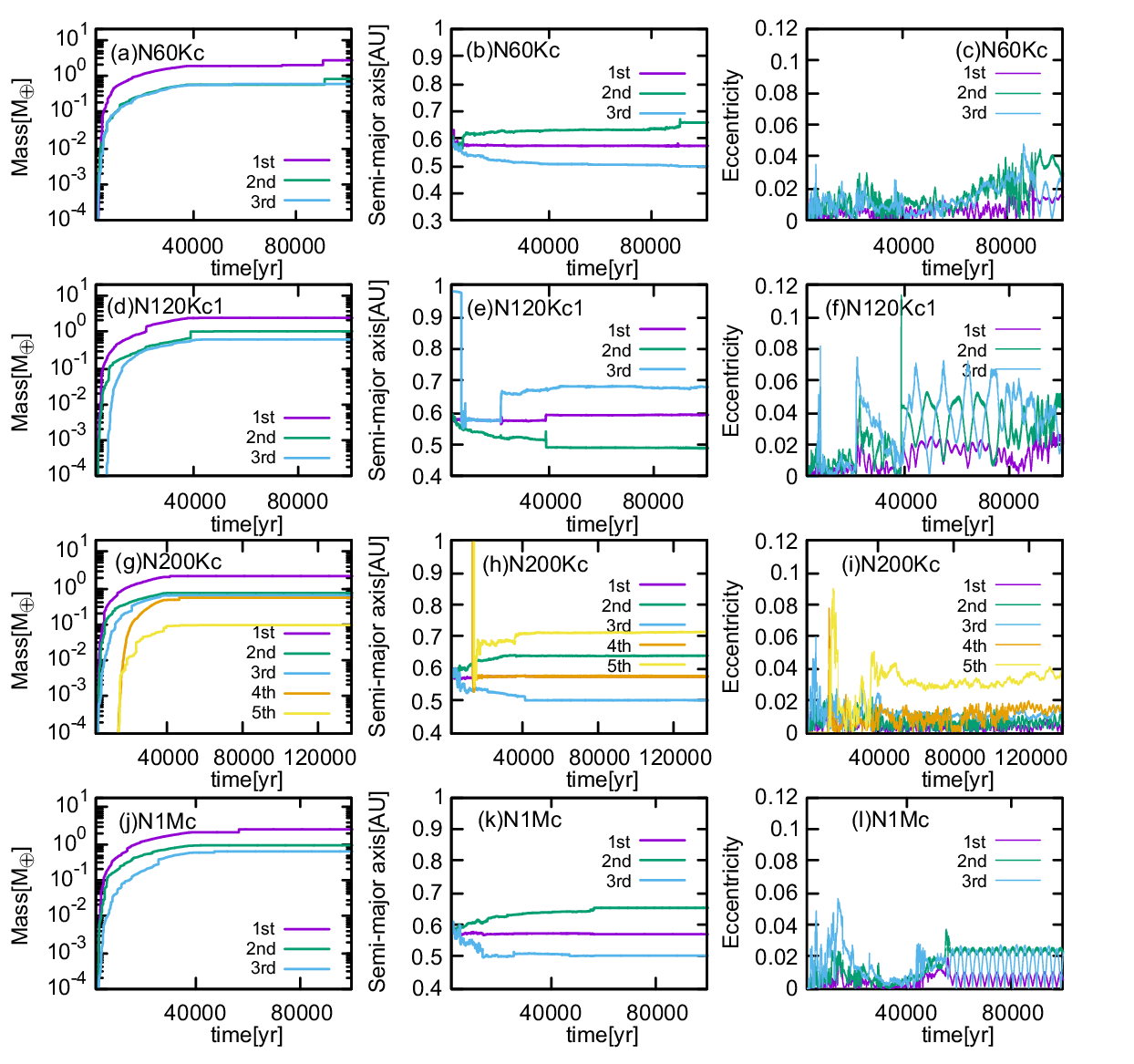}
 \end{center}
 \caption{The simulation results for the $6\times10^4$, $1.2\times10^5$, $2\times10^5$, and $1\times10^6$ particle systems in order from the top row to the bottom row. The left, middle and right surfaces are the results of the time evolution of the particle mass, semi-major axis, and eccentricity, respectively.
 }
 \label{fig:particle_number}
\end{figure*}

Fig.\ref{fig:particle_number} shows the evolution of the most massive plants (mass, semi-major axis, and eccentricity) for runs N60Kc to N1Mc, where we changed the total number of particles while keeping other parameters the same.
We can see that the evolution is very similar for runs with widely different numbers of particles. The mass of the most massive planet is around $2~M_{\oplus}$ for all runs, and that of the second and third massive planets are around $1~M_{\oplus}$ and $0.6~M_{\oplus}$. The semi-major axes are around 0.58, 0.5, and 0.65 AU for the most massive, second (or third), and third (or second) most massive planets. Final eccentricities are less than 0.05 for all cases. 

Since dust accumulates most at IDB where the pressure maximum is, it is typically considered that planets form most efficiently at IDB. Subsequently formed planets and surrounding planetesimals are scattered by the most massive planet and distributed in and out of the boundary. \citet{2022MNRAS.510.5486C} also studied planet formation in the pebble ring located in IDB, and their study also found that the most massive planet tended to form in the center of the ring compared to the location of the second massive planet.

We can conclude that our result does not depend on the mass of
``pebbles'', even though our ``pebbles'' are still many orders of
magnitude more massive than real ones.

\section{Discussion}\label{sec:Discussion}
In this section, we discuss the relation between our model and those in recent studies in which the radial structure of the protoplanetary disk is taken into account.

\citet{2018A&A...612L...5O} considered both MRI-active and MRI-inactive disk structures in which the surface density is flat, or increasing outward a certain radius. In such disk profiles, the Type I
migration is suppressed.  In such a disk, it is possible that planet formation occurs in a wide area around 1 AU \citep{2018A&A...612L...5O}. They have simulated the late stage of planet formation from planetary embryos. They used the same initial condition as \citet{2009ApJ...703.1131H} assuming a uniform solid surface density ranging from $0.7 - 1$ AU with 400 bodies of equal mass. Each body has a mass of $0.005~M_{\oplus}$, resulting in a total mass of $2~M_{\oplus}$. They showed that Earth-like planets are formed through collisions of planetary embryos, with a formation timescale of 100 Myr.

In contrast to  \citet{2018A&A...612L...5O}, we conducted $N$-body simulation to directly simulate the formation and growth of planetary embryos through pebble accretion. The notable difference between our results and \citet{2018A&A...612L...5O} is the planet formation timescale. In our simulation, Earth-like planets are formed within 0.1 Myr. This is because, in our model, the continuous supply of pebbles from the outer disk allows the efficient growth of planetesimals through collisions with both planetesimals and pebbles. 

\citet{2023MNRAS.518.3877J} studied the formation and evolution of planetesimals in a clumpy ring and a ring induced by a permanent pressure bump by conducting $N$-body simulations. \citet{2022A&A...668A.170L} also investigated planetesimals growth and their migration within a pressure bump. In both studies, they focused on planet formation at a pressure bump located at a distance of 75 AU from the central star motivated by observations (e.g., \cite{2018ApJ...869L..41A}). In addition, \citet{2022A&A...668A.170L} studied the planet formation process at a pressure bump located at a distance of 10 AU from the central star. In our study, we are interested in the planet formation process around 1 AU. Thus, it is difficult to compare the result of their work and our work directly.

In our study, planetesimals formed early on the high-mass end of the cumulative mass distribution grow rapidly, leaving out other less massive bodies by pebble accretion (Figs. \ref{fig:cumulative_mass}).  These few massive bodies continue to grow efficiently until the pebble infall phase ends. During the pebble infall phase, the total number of particles in the vicinity of IDB remains nearly constant, despite the continuous supply of dust particles. This indicates that few massive bodies eat the infalling particles while stirring a certain number of small bodies and preventing the growth of those small ones left behind. After the end of the pebble infall phase, these few massive bodies eat up all remaining small bodies, and only a small number of massive bodies are formed. The orbital separations between these few massive bodies are more than  $5~r_{\mathrm{H}}$ at the end of our simulations (Fig.\ref{fig:semi_major_comparison}). This is because they undergo scattering between planets \citep{1995Icar..114..247K}.
 
\section{Summary and Conclusion}\label{sec:Summary}
In our study, we simulated the formation of a solar-system-like terrestrial planetary system with pebble accretion and particle coalescence by performing \textit{N}-body simulations with a large number of particles ($6\times 10^4$ to $1\times 10^6$ particles).

We started from a protoplanetary disk inferred from recent standard understanding, which has dead zone in the radial range of around 0.5-3 AU. We followed the growth and inward migration of dust in this region and study how they form planets at the inner  dead zone boundary (IDB) through $N$-body simulation. Our findings are summarized as follows.

\begin{enumerate}
\item We found that planet formation at IDB is quite efficient. Independent of the inflow rate of the pebbles, massive planets are formed by the end of the pebble accretion phase. Runway growth leads to the formation of three to four massive planets, and their total mass becomes comparable to the total mass inflow rather early.

\item Our results do not depend on the mass resolution of our $N$-body simulation. Therefore, we can expect that a similar formation process take place even for real pebbles with much smaller mass, as far as the total inflow mass is similar.

\item Since the planets grow mainly through pebble accretion, their final eccentricities are relatively small. Even so, there are several collision events that can be regarded as giant impacts. 
\end{enumerate}\par
Based on our findings, we can conclude that there is a possibility for the formation of planetary systems resembling the Solar System within radially structured protoplanetary disks. Recent observational studies using the Atacama Large Millimeter/sub-millimeter Array (ALMA) have revealed the presence of dust ring-gap structures even within young protoplanetary disks, approximately the age of 1 Myr after the central star has formed (e.g., \cite{2018ApJ...869L..46D}).  A more recent study has suggested that planetary system formation proceeds rapidly with the timescale of 0.1 - 1 Myr after the central star has formed (\cite{2023ApJ...951....8O}). Our results may provide theoretical support for the efficient formation of planetary systems within such dust rings.

The main limitation of our current model is that the structure of the gas disk is oversimplified. We essentially placed a sharp cutoff of the gas disk at IDB. As suggested by many recent works (e.g. \authorcite{2016A&A...596A..74S} \yearcite{2010ApJ...718.1289S}; \yearcite{2016A&A...596A..74S}), the inner boundary of the gas disk might not be  a sharp one but a more smooth transition driven by magnetic disk wind. It is also important to take into account the effect of the Type-I migration  \citep{2023MNRAS.518.3877J, 2022A&A...664A..86J}. We will extend our work to study these effects.
%%%%%%%%%%%%%%%%%%%%%%%%%%%%%%%%%%%%%%%

\begin{ack}
We thank the anonymous referee for providing useful feedback that contributed to the improvement of our paper. This work was supported by MEXT as ``Program for Promoting Researches on the Supercomputer Fugaku" (Structure and Evolution of the Universe Unraveled by Fusion of Simulation and AI; Grant Number JPMXP1020230406) and used computational resources of supercomputer Fugaku provided by the RIKEN Center for Computational Science (Progect ID: hp230204).
The simulations in this paper were also carried out on a Cray XC50 system at the Centre for Computational Astrophysics (CfCA) of the National Astronomical Observatory of Japan (NAOJ). 
\end{ack}

\appendix 
\section{Mass flux onto IDB in our $N$-body simulation}\label{sec:Appendix_mass_flux}
In our model, we assumed that dust particles grow in situ and begin to drift toward the Sun once their mass reaches $m_{\mathrm{c}}$. This means that the accretion time to IDB from the beginning of the simulation, $t_{\mathrm{acc}}$, consists of the sum of the two timescales:
\begin{equation}
t_{\mathrm{acc}}(r)=t_{\mathrm{grow}}(r)+t_{\mathrm{drift}}(r), \label{eq.41}
\end{equation}
where $r$ is the initial position of a dust particle, $t_{\mathrm{grow}}$ is the time it takes for a dust particle to grow to $m_{\mathrm{c}}$ and $t_{\mathrm{drift}}$ is the particle drift timescale given by eq.(\ref{eq.40}).
% %Figure. 5
\begin{figure}[hbtp]
 \includegraphics[width= 8cm]{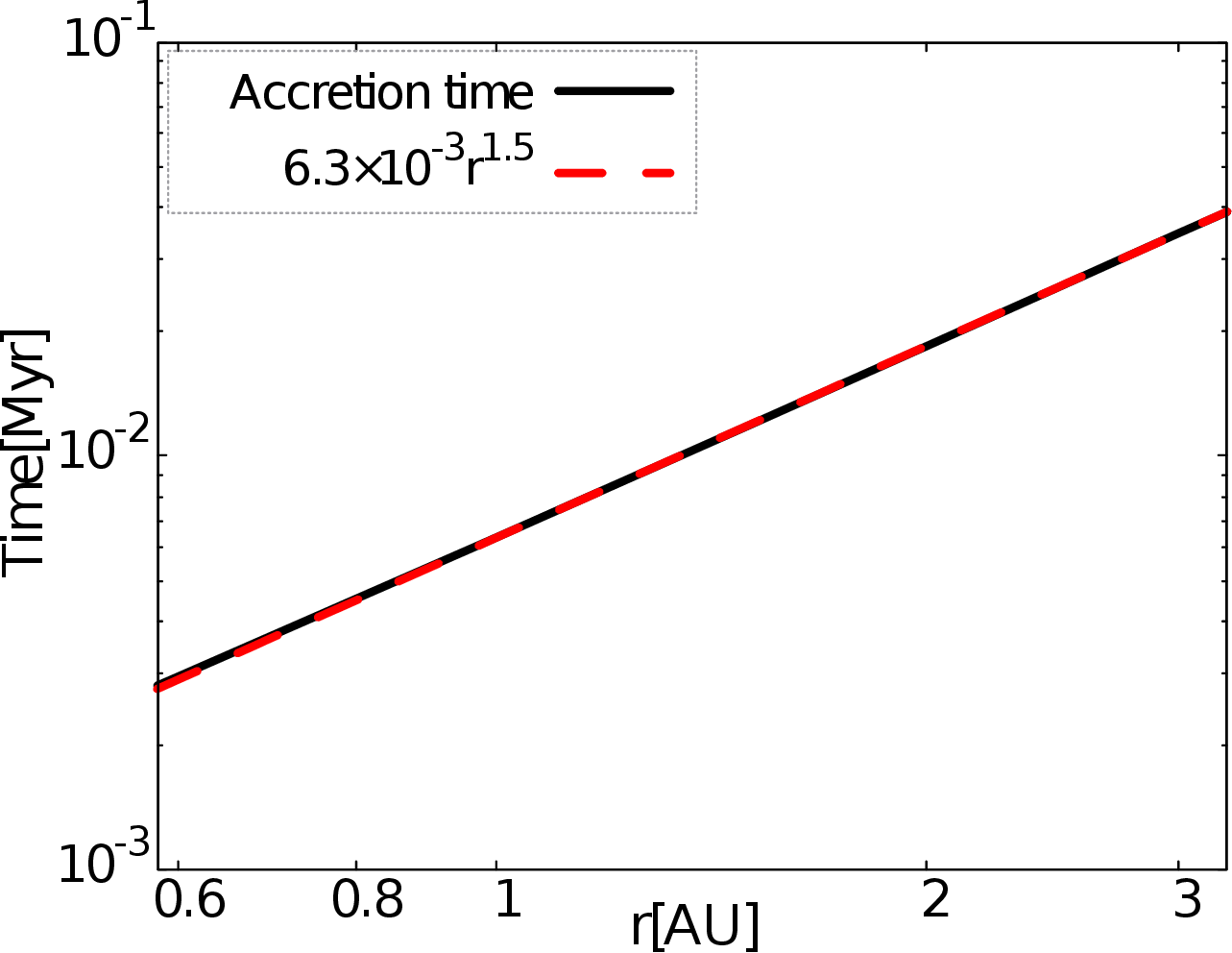}
 \caption{The accretion time $t_{\mathrm{acc}}$ at different radii in the disk (black line).\\ The approximation of $t_{\mathrm{acc}}$ as a function of the disk radius (red dashed line). The accretion rate and the dust-to-gas fraction are $\dot{M}=10^{-7.0}~M_{\odot}/\mathrm{yr}$ and $\bar{f}=\bar{f}_{\mathrm{MMSN}}$, respectively.} 
 \label{fig:acc_time}
\end{figure}\par
Fig.\ref{fig:acc_time} shows the accretion time as a function of the dust particle position $r$. We draw a fitting line with a simple power law form (red dashed line) in the figure as well as the numerical simulation (black line). The function form of the accretion timescale is
 \begin{equation}
 t_{\mathrm{acc}}(r)=6.3\times10^{-3}r^{1.5}~~~\mathrm{Myr}. \label{eq.42}
 \end{equation}
By solving this equation for $r$, we can obtain the distance where the dust particles just reach IDB at a given time. \par

In our $N$-body simulations, we reproduced the accretion of dust particles onto IDB in accordance with $t_{\mathrm{acc}}$ given by eq.(\ref{eq.42}).
To verify that our $N$-body simulation accurately reproduces the mass flux, we need to compare the analytically derived mass flux using $t_\mathrm{acc}$ with our $N$-body simulation result. In the following, we briefly describe a method for determining the mass flux analytically and compare it with our $N$-body simulation result. \par
%Figure: dust surface density
\begin{figure}[hbtp]
 \includegraphics[width= 8cm]{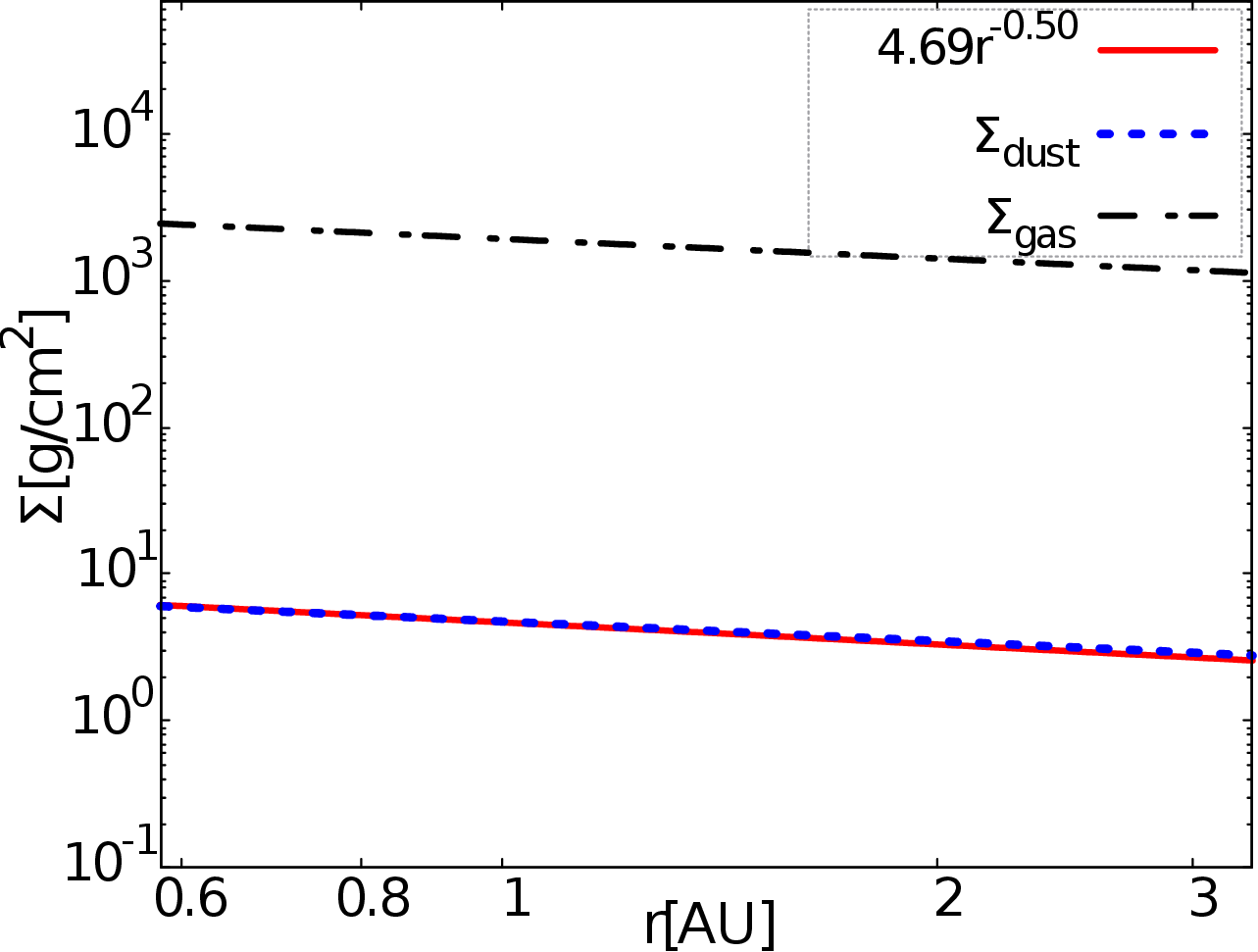}
 \caption{ The radial profile of the surface density of the dust (blue dotted line) and gas (black dash-dotted line) for $\dot{M}=10^{-7.0}~M_{\odot}/$yr ranging from $r_{\mathrm{in}}$ to $r_{\mathrm{WSZ}}$. The approximation of $\Sigma_{\mathrm{dust}}$ as a function of the disk radius (red line).} 
 \label{fig:dust_surface_density}
\end{figure}
Fig.\ref{fig:dust_surface_density} shows the result of $\Sigma_{\mathrm{dust}}$ and $\Sigma_{\mathrm{gas}}$ for $\dot{M}=10^{-7.0}~M_{\odot}/$ yr ranging from $r_{\mathrm{in}}=0.58$ AU to $r_{\mathrm{WSZ}}=3.28$ AU. Here we note that $r_{\mathrm{in}}$ and $r_{\mathrm{WSZ}}$ represent the inner boundaries of the dead zone and water sublimation zone, respectively. According to  Fig.\ref{fig:dust_surface_density}, the dust surface density $\Sigma_{\mathrm{dust}}$ in MSZ can be approximated as a power function of distance $r$, yielding the following functional form:
 \begin{equation}
\Sigma_{\mathrm{dust}}=\bar{f}\Sigma_{\mathrm{gas}}=4.69 \left(\frac{\bar{f}}{\bar{f}_{\mathrm{MMSN}}}\right)r^{-0.5}~~~\mathrm{g}~\mathrm{cm}^{-2}. \label{eq.43}
\end{equation}
Solving eq.(\ref{eq.42}) for $r$, we can determine how far dust particles accrete at a given time (here we denote $r$ as $r_{\mathrm{acc}}$). By substituting $r_\mathrm{acc}$ into eq.(\ref{eq.43}) and integrating it over time, we can calculate how much dust has accreted onto IDB by a certain time. Here, we call it ``theoretical mass flux".\par
%Figure. 6
\begin{figure}[hbtp]
 \includegraphics[width= 8cm]{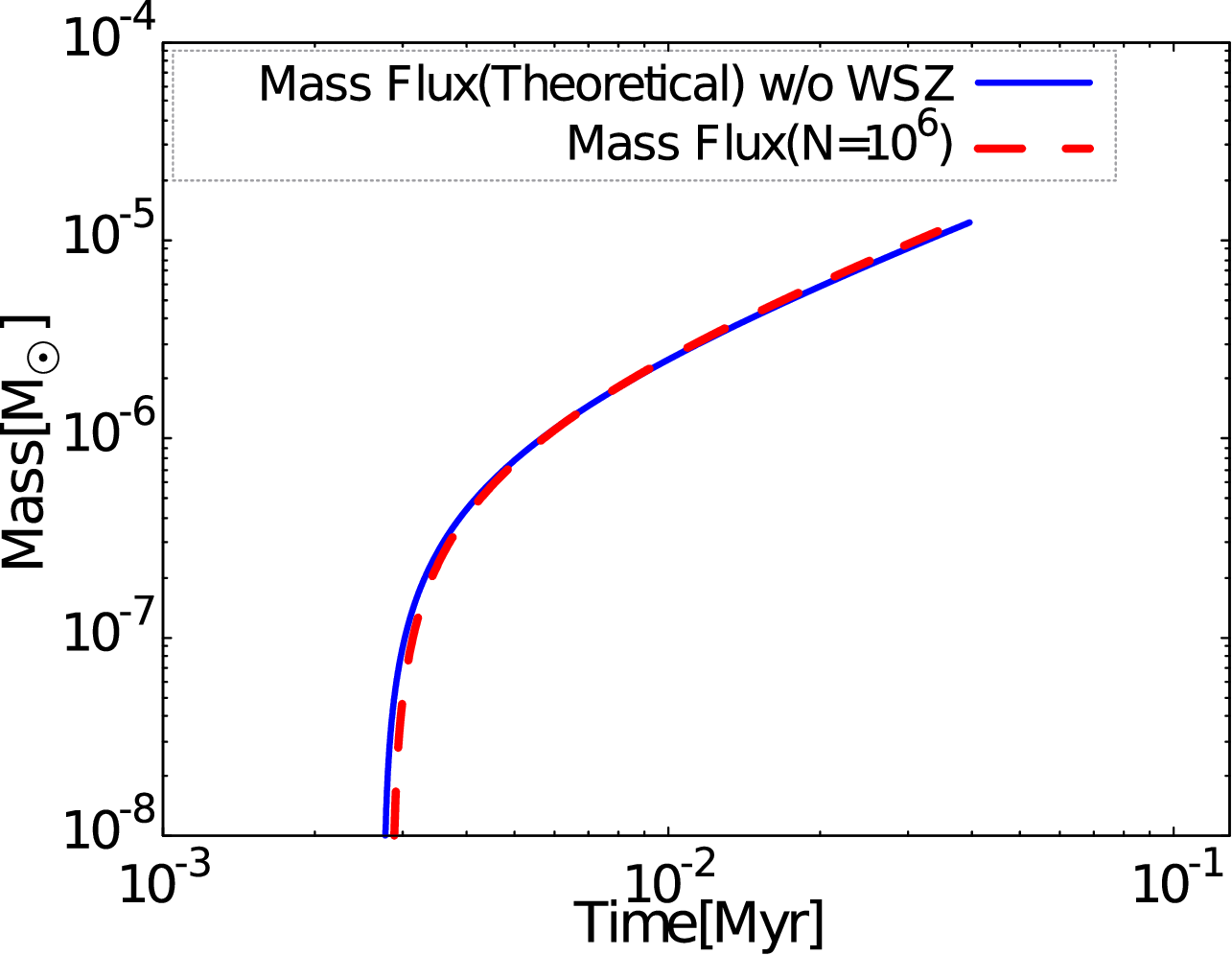}
 \caption{ Comparison of theoretical mass flux with the \textit{N}-body simulation result.  The vertical axis represents the total mass flux onto IDB up to a certain time. The theoretical values (blue line) and the model for $1\times10^6$ particles (red dashed line).} 
 \label{fig:mass_flux}
\end{figure}
In Fig.\ref{fig:mass_flux}, we show the results for the theoretical mass flux (blue line) and the model for $1\times10^6$ particles (red dashed line)  with $\bar{f}=\bar{f}_{\mathrm{MMSN}}$ disk. Fig.\ref{fig:mass_flux} supports that our \textit{N}-body simulations successfully reproduce mass inflow into IDB with high accuracy.

\section{Energy error in our \textit{N}-body simulation} \label{sec:Appendix1}
In our model, there are sudden mass changes in particles, resulting in non-mechanical energy changes. It is, thus, necessary to count these energy changes to see the total energy errors in our simulations. In our simulation code, the change of mass takes place at the end of the hard part when the particle passes through IDB or collides with another particle. By keeping the energies before the mass change, we can count the total energy change induced by this operation.

The energy change due to the mass changes of particles, i.e., the non-mechanical energy change, is calculated as the summation of the increase of the kinetic energy and gravitational energy of the particle due to the change in mass. 
The non-mechanical energy change at a step is as follows:

\begin{eqnarray}
\Delta E_{\mathrm{n.m.}}&=& \sum_i \lbrack\epsilon_{0,i}+\epsilon_{1,i}+\epsilon_{2,i}+\epsilon_{3,i}\rbrack, \label{eq:dEnm}\\
\epsilon_{0,i}&=&\frac{1}{2}\Delta m_{\mathrm{i}}{v}_{\mathrm{p}}^2,\\
\epsilon_{1,i}&=&-G\Delta m_{i}\sum_{k\in  N_{i}} m_{k}\left\lbrack\frac{W(r_{ij};r_{\mathrm{out},ij})}{r_{ik}}\right\rbrack \\
\epsilon_{2,i}&=&-\frac{G \Delta m_{\mathrm{i}}M_{*}} {r_{i}},\\
\epsilon_{3,i}&=&\Delta m_{i}\phi_{\mathrm{soft},i},
\end{eqnarray}
where $\Delta m_{i}$ is the change in mass of $i$-th particle ($m_{\mathrm{new}} = m_{\mathrm{old}} + \Delta m_{i}$), $W(r_{ij};r_{\mathrm{out},ij})$ is the cutoff function used in P$^3$T, $N_i$ is the number of neighbor particles of the $i$-th particle, $\phi_{\mathrm{soft},i}$ is the soft part of the gravitational potential for $i$-th particle. 
The four terms of eq. (\ref{eq:dEnm}) represent the energy changes of the kinetic energy, the gravitational potential energy in the short-range interactions, the gravitational potential energy with respect to the central star, and the gravitational potential energy in the long-range interactions, respectively.
 %Figure. A1
\begin{figure}[hbtp]
 %\centering
 \includegraphics[width= 8cm]{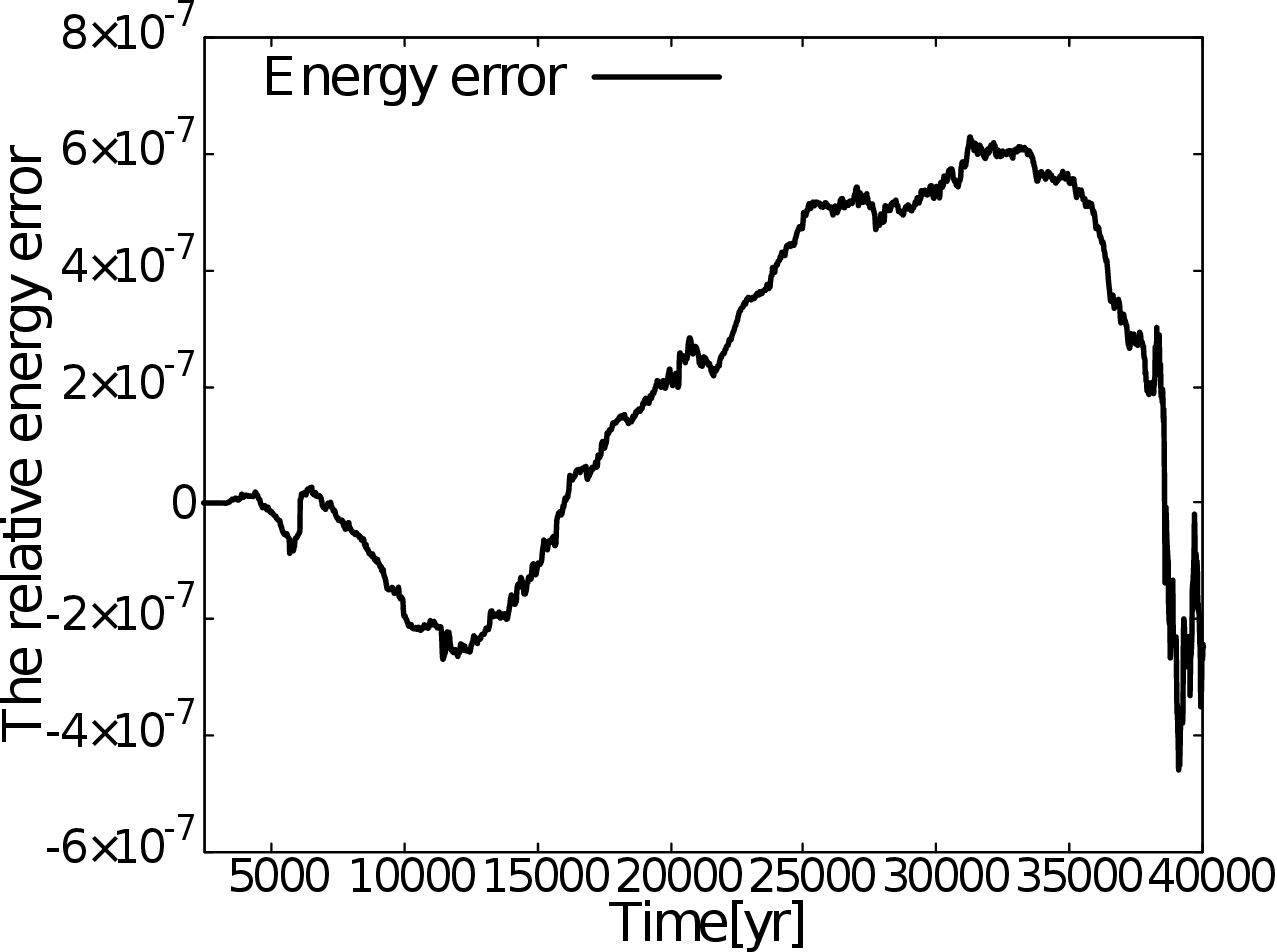}
 \caption{The time evolution of the relative energy error for N120Kc1. The relative energy error is given as $(E_{\mathrm{now}}-E_{\mathrm{initial}}-E_{\mathrm{n.m.}})/E_{\mathrm{now}}$.} 
 \label{fig.A1}
\end{figure}
Fig.\ref{fig.A1} shows the relative energy error of N120Kc1 involving the energy correction due to mass changes as a function of time. From this figure, we can confirm that the correction of the sudden mass changes functioned properly. Note that Fig.\ref{fig.A1} shows only the first $4\times10^4~{\rm yr}$ since the accretion stops around the epoch and there is no need to correct induced by the mass changes after this epoch.

\bibliographystyle{aasjournal}
\bibliography{reference}
\end{document}